\documentclass[letterpaper,UKenglish]{dagman}

\title{Research Directions for Principles of
Data Management\\(Dagstuhl Perspectives Workshop
16151)}
\author{Serge Abiteboul}
\author{Marcelo Arenas}
\author{Pablo Barcel\'o}
\author{Meghyn Bienvenu}
\author{Diego Calvanese}
\author{Claire David}
\author{Richard Hull}
\author{Eyke H\"ullermeier}
\author{Benny Kimelfeld}
\author{Leonid Libkin}
\author{Wim Martens}
\author{Tova Milo}
\author{Filip Murlak}
\author{Frank Neven}
\author{Magdalena Ortiz}
\author{Thomas Schwentick}
\author{Julia Stoyanovich}
\author{Jianwen Su}
\author{Dan Suciu}
\author{Victor Vianu}
\author{Ke Yi}

\authorrunning{S.\ Abiteboul et al.}
\titlerunning{Research directions for Principles of
Data Management}

\begin{document}

\def\e#1{\emph{#1}}
\def\partitle#1{\par\vskip1em\noindent\textbf{#1.}\,\,}

\maketitle

\makeatletter{}\section{Introduction}

In April 2016, a community of researchers working in the area of
Principles of Data Management (PDM) joined in a workshop at the
Dagstuhl Castle in Germany.  The workshop was organized jointly by the
Executive Committee of the ACM Symposium on Principles of Database
Systems (PODS) and the Council of the International Conference on
Database Theory (ICDT).  The mission of this workshop was to identify
and explore some of the most important research directions that have
high relevance to society and to Computer Science today, and where the
PDM community has the potential to make significant contributions.
This report describes the family of research directions that the
workshop focused on from three perspectives: potential practical
relevance, results already obtained, and research questions that
appear surmountable in the short and medium term.
This report organizes 
the identified research challenges for PDM around seven
core themes, namely 
  {\em Managing Data at Scale}, 
  {\em Multi-model Data}, 
  {\em Uncertain Information}, 
  {\em Knowledge-enriched Data}, 
  {\em Data Management and Machine Learning}, 
  {\em Process and Data}, and 
  {\em Ethics and Data Management}.
Since new challenges in PDM arise all the time, we note
that this list of themes is not intended to be exclusive.  

This report is intended for a diverse audience.
It is intended for government and industry funding agencies,
because it includes an articulation of important areas where the PDM
community is already contributing to the key data management
challenges in our era, and has the potential to contribute much more.
It is intended for universities and colleges world-wide,
because it articulates the importance of continued research
and education in the foundational elements of data management,
and it highlights growth areas for Computer Science and
Management of Information Science research.
It is intended for researchers and students, because it identifies
emerging, exciting research challenges in
the PDM area, all of which have very timely practical relevance.
It is also intended for policy makers, sociologists, and philosophers,
because it re-iterates the importance of considering ethics
in many aspects of data creation, access, and usage, and
suggests how research can help to find
new ways for maximizing the benefits of massive data while
nevertheless safeguarding the privacy and integrity of
citizens and societies. 

The field of PDM is broad.
It has ranged from the development of formal frameworks for 
understanding and managing
data and knowledge (including data models, query languages, ontologies,
and transaction models)
to data structures and algorithms (including query optimizations,
data exchange mechanisms, and privacy-preserving manipulations).
Data management is at the heart of most IT applications today, and
will be a driving force in personal life, social life, industry, and
research for the foreseeable future.
We anticipate on-going expansion of
PDM research as the
technology and applications involving data management
continue to grow and evolve.

PDM played a foundational role in the relational database model,
with the robust connection between algebraic and calculus-based query 
languages,
the connection between integrity constraints and database design,
key insights for the field of query optimization,
and the fundamentals of consistent concurrent transactions.
This early work included rich cross-fertilization between
PDM and other disciplines in mathematics and computer science,
including logic, 
complexity theory,
and knowledge representation.
Since the 1990s we have seen
an overwhelming increase in both the production of data 
and the ability to store and access such data.
This has led to a phenomenal metamorphosis in
the ways that we manage and use data.  
During this time, we have gone
(1) from stand-alone disk-based databases to data that is spread
across and linked by the Web, (2) from rigidly structured towards
loosely structured data, and (3) from relational data to many
different data models (hierarchical, graph-structured, data points,
NoSQL, text data, image data, etc.). 
Research on PDM has
developed during that time, too, following, accompanying and
influencing this process. It has intensified research on extensions 
of the relational model (data exchange, incomplete data, probabilistic
data, \dots), 
on other data models (hierachical, semi-structured, graph, text, \dots),
and on a variety of further data management areas, including 
knowledge representation and
the semantic web, 
data privacy and security,
and data-aware (business) processes. 
Along the way,
the PDM community expanded its cross-fertilization
with related areas, to include
automata theory,
web services,
parallel computation,
document processing, 
data structures, scientific workflow, business process management, 
data-centered dynamic systems, 
data mining, machine learning, information extraction, etc. 

Looking forward,
three broad areas of data management stand out where principled,
mathematical
thinking can bring new approaches and much-needed clarity. 
The first relates to the full lifecycle of so-called ``Big Data Analytics'',
that is, the application of statistical and machine
learning techniques to make sense out of, and derive value from,
massive volumes of data.
The second stems from new forms of data creation and processing, 
especially as it arises in applications such as
web-based commerce, social media applications, and 
data-aware workflow and business process management.
The third, which is just beginning to emerge, 
is the development of new principles and approaches
in support of ethical data management.
We briefly illustrate some of the primary ways
that these three areas can be supported by the
seven PDM research themes that are explored in this report.

The overall lifecycle of Big Data Analytics raises a wealth of
challenge areas that PDM can help with.  As documented in numerous
sources, so-called ``data wrangling'' can form 50\% to 80\% of the
labor costs in an analytics investigation.  The challenges of data
wrangling can be described in terms of the ``4 V's'' -- Volume,
Velocity, Variety, and Veracity -- all of which have been addressed,
and will continue to be addressed, using principled approaches.  As we
will discuss later, PDM is making new contributions towards managing
the Volume and Velocity, as discussed in
{\em Managing Data at Scale} (Section 2). For
example, there have been recent advances in efficient $n$-way join
processing in highly parallelized systems, which outperform
conventional approaches based on a series of binary joins
\cite{atserias2008size,chu15:_from}.  PDM is contributing towards managing the
Variety: {\em Knowledge-enriched Data} (Section 5) provides tools
for managing and efficient reasoning with industrial-sized ontologies
\cite{CGLLR07}, and {\em Multi-model Data} (Section 3) provides
approaches for efficient access to diverse styles of data, from
tabular to tree to graph to unstructured.  Veracity is an especially
important challenge when performing analytics over large volumes of
data, given the inevitability of inconsistent and incomplete data.
The PDM field of {\em Uncertain Information} (Section 4) has provided a formal
explanation of how to answer queries in the face of uncertainty some four
decades ago \cite{Lipski-certain-answers}, but its computational
complexity has made mainstream adoption elusive -- a challenge that
the PDM community should redouble its efforts to resolve.  
Provocative new opportunities are raised in the
area of {\em Data Management and Machine Learning} (Section 6),
because of the unconventional ways in which
feature engineering and machine learning 
algorithms access and manipulate large
data sets.  We are also seeing novel approaches to incorporate Machine Learning
techniques into database management systems, e.g., to enable more
efficient extraction and management of information coming from text
\cite{DBLP:conf/sigmod/ArefCGKOPVW15}.

The new forms of data creation and processing that have emerged
have led to new forms of data updates,
transactions, and data management in general.
Web-based commerce has revolutionized how business works with
supply chain, financial, manufacturing, and other kinds of data,
and also how businesses engage with their customers, both consumers and
other businesses.
Social applications have revolutionalized our personal and social
lives, and are now impacting the workplace in similar ways.
Transactions are increasingly distributed, customized, 
personalized, offered with more immediacy,
and informed by
rich sets of data and advanced analytics.
These trends are being compounded as the Internet of Things 
becomes increasingly real and leveraged to increase
personal convenience and business efficiencies.
A broad challenge is to make it easy to understand all of this data, and the ways that the data are being processed; approaches to this challenge are offered in both
{\em Multi-model Data} (Section 3) and
{\em Knowledge-enriched Data} (Section 5).
Many forms of data from the Web,
including from social media, from crowd-sourced query
answering, and unstructured data in general
create {\em Uncertain Information} (Section 4).
Web-based communication has also enabled a revolution
in electronically supported processes, ranging from
conventional business processes that are now becoming
partially automated, to consumer-facing e-commerce systems,
to increasingly streamlined 
commercial and supply chain applications.
Approaches have emerged
for understanding and managing {\em Process and Data} (Section 7) in a holistic manner,
enabling a new family of automated verification techniques
\cite{Calvanese-et-al-PODS-2013}; these will become increasingly important
as process automation accelerates.

While ethical use of data has always been a concern,
the new generation of data- and information-centric applications, 
including Big Data Analytics, social applications, and
also the increasing use of data in commerce 
(both business-to-consumer and business-to-business)
has made ethical considerations more important and more challenging.
At present there are huge volumes of data being collected about
individuals, and being interpreted in many different ways by 
increasing numbers of diverse organizations with widely
varying agendas.
Emerging research suggests that the use of mathematical principles
in research on {\em Ethics and Data Management} (Section 8)
can lead to new approaches to ensure data privacy for individuals, and
compliance with government and societal regulations
at the corporate level. As just one example, mechanisms are emerging 
to ensure accurate and ``fair'' representation of 
the underlying data when analytic techniques are applied
\cite{Detal12}.

The findings of this report differ from, and complement, the 
findings of the 2016 Beckman Report 
\cite{DBLP:journals/cacm/AbadiAABBCCDDFG16} in two
main aspects.
Both reports stress the importance of ``Big Data'' as
the single largest driving force in data management usage and research 
in the current era.
The current report focuses primarily on research challenges where 
a mathematically based 
perspective has had and will continue to have substantial impact. 
This includes for example new algorithms for large-scale parallelized 
query processing and Machine Learning, 
and models and languages for heterogeneous and
uncertain information.
The current report also considers additional areas where research
into the principles of data management can make growing
contributions in the coming years, including
for example approaches for combining data structured
according to  different models, 
process taken together with data, and ethics
in data management.

The remainder of this report includes the seven technical sections
mentioned above, and a concluding section with comments about the
road ahead for PDM research.

\makeatletter{}\section{Managing Data at Scale}\label{sec-mds}
Volume is still the most prominent feature of \emph{Big Data}.  The
PDM community, as well as the general theoretical computer
science community, has made significant contributions to efficient
data processing at scale.  This is evident from the tremendous success
of parallel algorithms, external memory algorithms, streaming
algorithms, etc., with their applications in large-scale database
systems.  Sometimes, the contributions of  theoretical foundations might
not be immediate,
e.g., it took more than a decade for the \emph{MapReduce}
system to popularize Valiant's theoretical \emph{bulk synchronous parallel
(BSP)} model \cite{valiant90} in the systems community.  But this exactly means that one should
never underestimate the value of theory. We face the following
practical challenges:

\begin{description}
\item[New Paradigms for Multi-way Join Processing.]
A celebrated result by Atserias, Grohe, and Marx
\cite{atserias2008size}  has sparked a flurry of research
efforts in re-examining how multi-way joins should be computed.  In all current
relational database systems, a multi-way join is processed in a pairwise framework
using a binary tree (plan), which is chosen by the query optimizer.  However, the
recent theoretical studies have discovered that for many queries and data instances,
even the best binary plan is suboptimal by a large polynomial factor.  Meanwhile,
worst-case optimal algorithms have been designed in the RAM model \cite{ngo2012worst},
the external memory model \cite{hu16:_towar}, and BSP models
\cite{beame13:_commun,afrati11:_optim}. 
 These new algorithms have all
abandoned the binary tree paradigm, while adopting a more \emph{holistic} approach to
achieve optimality.  Encouragingly, there have been empirical studies
\cite{chu15:_from} that demonstrate the practicality of these new algorithms. In
particular, \emph{leapfrog join} \cite{veldhuizen14}, a worst-case optimal algorithm, has
been implemented inside a full-fledged database system.  Therefore, we believe that the
newly developed algorithms in the theory community have a potential to change how
multi-way join processing is currently done in database systems.  Of course, this can
only be achieved with significant engineering efforts, especially in designing and
implementing new query optimizers and cost estimation under the new paradigm.
 
\item[Approximate query processing.]
 Most analytical queries on \emph{Big Data} return aggregated answers that
do not have to be 100\% accurate.  The line of work on {\em online
  aggregation} \cite{hellerstein97:_onlin} studies new algorithms that
allow the query processor to return approximate results (with
statistical guarantees) at early stages of the processing so that the
user can terminate it as soon as the accuracy is acceptable.  This
both improves interactiveness and reduces unnecessary resource
consumption.  Recent studies have shown some encouraging results
\cite{haas99:_rippl,li16:_wander}, but there is still a lot of room
for improvement: (1) The existing algorithms have only used simple
random sampling or sample random walks to sample from the full query
results.  More sophisticated techniques based on Markov Chain Monte
Carlo might be more effective.  (2) The streaming algorithms community
has developed many techniques to summarize large data sets into
compact data structures while preserving important properties of the
data.  These data summarization techniques can be useful in
approximate query processing as well.  (3) Actually integrating these
techniques into modern data processing engines is still a significant
practical challenge.
\end{description}

These practical challenges raise the
following theoretical challenges:

\begin{description}
\item[The Relationship Among Various Big Data Computation Models.]
The theoretical computer science community has developed many
beautiful models of computation aimed at handling data sets that are
too large for the traditional random access machine (RAM) model, the most prominent ones
including parallel RAM (PRAM), external memory (EM) model, streaming model, the BSP
model and its recent refinements to model modern distributed
architectures.  Several studies seem to suggest that there are deep
connections between seemingly unrelated Big Data computation
models for streaming computation, parallel processing, and external
memory, especially for the class of problems interesting to the
PDM community (e.g., relational algebra) \cite{ST94,FMSSS08,KBS16}.  Investigating
this relationship would reveal the inherent nature of these problems
with respect to scalable computation, and would also allow us to
leverage the rich set of ideas and tools that the theory community
has developed over the decades.

\item[The Communication Complexity of Parallel Query Processing.]
New large-scale data analytics systems use massive parallelism to support complex
queries on large data sets.  These systems use clusters of servers and proceed in
multiple communication rounds.  In these systems, the communication cost is usually the
bottleneck, and therefore has become the primary measure of complexity for algorithms
designed for these models.  Recent studies
(e.g., \cite{beame13:_commun}) 
have
established tight upper and lower bounds on the communication cost for computing some
join queries, but many questions remain open: (1) The existing bounds are tight only
for one-round algorithms.  However, new large-scale systems like Spark have greatly
improved the efficiency of multi-round iterative computation, thus the one-round limit
seems unnecessary.  The communication complexity of multi-round computation remains
largely open.  (2) The existing work has only focused on a small set of queries (full
conjunctive queries), while many other types of queries remain unaddressed. Broadly,
there is great interest in large-scale machine learning using these systems, thus it is
both interesting and important to study the communication complexity of classical
machine learning tasks under these models. 
This is developed in more detail in Section \ref{sec-dmml}, which summarizes research opportunites 
at the crossroads of data management and machine learning. Large-scale
parallel query processing raises many other (practical and
foundational) research questions. As an example, 
recent frameworks for parallel query optimization 
 need to be extended to the multi-round case \cite{AmelootGKNS15}.

\end{description}

We envision that the following theory techniques will be useful in addressing the
challenges above (that are not considered as ``classical'' PDM or database theory):
Statistics, sampling theory, approximation theory, communication complexity,
information theory, convex optimization.

\makeatletter{}\section{Multi-model Data: Towards an Open Ecosystem of Data Models}\label{sec-mmd}

Over the past 20 years, the landscape of available data has
dramatically changed. While the huge amount of available data is
perceived as a clear asset, exploiting this data meets the challenges of the ``4 V's'' mentioned in the Introduction. 

One particular aspect of the \emph{variety} of data is the existence and
coexistence of different models for semi-structured and unstructured
data, in addition to the widely used relational data model. Examples
include tree-structured data (XML, JSON), graph data (RDF, property
graphs, networks), tabular data (CSV), temporal and spatial data, text, and multimedia. We can expect that in the near future, new data models will arise in order to cover particular needs.
Importantly, data models include not only a data structuring
paradigm, but also approaches for queries,
updates, integrity constraints,
views, integration, and transformation, among others.

Following the success of the relational data model, originating from the close interaction between theory and practice, the PDM community has been working for many years towards understanding each one of the aforementioned models formally. Classical DB topics---schema and query languages, query evaluation and optimization, incremental processing of evolving data, dealing with inconsistency and incompleteness, data integration and exchange, etc.---have been revisited. This line of work has been successful from both the theoretical and practical points of view. As these questions are not yet fully answered for the existing data models and will be asked again whenever new models arise, it will continue to offer practically relevant theoretical challenges.  But what we view as a new grand challenge is the coexistence and interconnection of all these models, complicated further by the need to be prepared to embrace new models at any time. 

The coexistence of different data models resembles the fundamental problem of data heterogeneity within the relational model, which arises when semantically related data is organized under different schemas. This problem has been tackled by data integration and data exchange, but since these classical solutions have been proposed, the nature of available data has changed dramatically, making the questions open again. This is particularly evident in the Web scenario, where not only the data comes in huge amounts, in different formats, is distributed, and changes constantly, but also it comes with very little information about its structure and almost no control of the sources. Thus, while the existence and coexistence of various data models is not new, the recent changes in the nature of available data raise a strong need for a new principled approach for dealing with different data models: an approach flexible enough to allow keeping the data in their original format (and be open for new formats), while still providing a convenient unique interface to handle data from different sources. It faces the following four specific practical challenges.

\begin{description}
\item[Modelling data.] How does one turn raw data into a database? This used to amount to designing the right structure within the relational model. Nowadays, one has to first choose the right data models and design interactions between them. Could we go even further and create methodologies allowing engineers to design a new data model? 
\item[Understanding data.] How does one make sense of the data? Previously, one could consult the structural information provided with the data. But presently data hardly ever comes with sufficient structural information, and one has to discover its structure. Could we help the user and systems to
understand the data without first discovering its structure in full?
\item[Accessing data.] How does one extract information? For years this meant writing an SQL query. Currently the plethora of query languages is perplexing and each emerging data model brings new ones. How can we help users formulate queries in a more uniform way?
\item[Processing data.] How does one evaluate queries efficiently? Decades of effort brought refined methods to speed up processing of relational data; achieving similar efficiency for other data models, even the most mature ones such as XML, is still a challenge. But it is time to start thinking about processing data combining multiple models (possibly distributed and incomplete).
\end{description}

These practical challenges raise concrete theoretical problems, some of which go beyond the traditional scope of PDM. Within PDM, the key theoretical challenges are the following. 

\begin{description}
\item[Schema languages.]
Design flexible and robust multi-model schema languages. Schema languages for XML and RDF data are standardized, efforts are being made to create standards for JSON \cite{DBLP:conf/www/PezoaRSUV16}, general graph data \cite{DBLP:conf/icdt/StaworkoBGHPS15}, and tabular data \cite{DBLP:conf/www/MartensNV15,AMRV16}. Multi-model schema languages should offer a uniform treatment of different models, the ability to describe mappings between models (implementing different views on the same data, in the spirit of data integration), and the flexibility to seamlessly incorporate new models as they emerge.
\item[Schema extraction.] Provide efficient algorithms to extract schemas from the data, or at least discover partial structural information (cf.~\cite{DBLP:journals/tods/BexNSV10,DBLP:conf/webdb/CafarellaSE07}). 
The long-standing challenge of entity resolution 
is exacerbated in the context of finding correspondences between
data sets structured according to different models
\cite{DBLP:journals/tods/CateDK13}. 
\item[Visualization of data and metadata.] 
Develop user-friendly paradigms for presenting the metadata information and statistical properties of the data in a way that helps in formulating queries. In an ideal solution, users would be presented relevant information about data and metadata as they type the query. This requires understanding and defining what the relevant information in a given context is, and representing it in a way allowing efficient updates as the context changes (cf. \cite{DBLP:journals/pvldb/CebiricGM15,DBLP:journals/ws/ArenasGKMZ16}).
\item[Query languages.]
Go beyond bespoke query languages for the specific data models~\cite{DBLP:conf/pods/ArenasGP14} and design a query language suitable for multi-model data, either incorporating the specialized query languages as sub-languages or offering a uniform approach to querying, possibly at the cost of reduced expressive power or higher complexity. 
\item[Evaluation and Optimization.]
Provide efficient algorithms for computing meaningful answers to a query, based on structural information about data, both inter-model and intra-model; this can be tackled either directly \cite{DBLP:conf/icdt/KaminskiK16,DBLP:journals/tods/GottlobKP05} or via static optimization \cite{DBLP:journals/jacm/BenediktFG08,DBLP:conf/pods/CzerwinskiMPP15}. In the context of distributed or incomplete information, even formalizing the notion of a meaningful answer is a challenge  \cite{DBLP:journals/ai/Libkin16}, as discussed in more detail in Section \ref{sec-ui}.
\end{description}

All these problems require strong tools from PDM and theoretical computer science in general (complexity, logic, automata, etc.). But solving them will also involve  knowledge and techniques from neighboring communities. For example, 
the second, third and fifth challenges naturally involve data mining and machine learning aspects (see Section \ref{sec-dmml}).
The first, second, and third raise knowledge representation issues (see Section~\ref{sec-kedm}). 
The first and fourth will require expertise in programming languages. 
The fifth is at the interface between PDM and algorithms, but also between PDM and systems. The third raises human-computer interaction issues.

\makeatletter{}
\def\benny#1{{\color{blue}\textbf{Benny:} #1}}
\def\bennynew#1{{\color{red}#1}}

\section{Uncertain Information}\label{sec-ui}
Incomplete, uncertain, and inconsistent information is ubiquitous in
data management applications. This was recognized already in the 1970s
\cite{codd75}, and since then the significance of the issues related
to incompleteness and uncertainty has been steadily growing: it is a fact
of life that data we need to handle on an everyday basis is rarely
complete.  However, while the data management field developed
techniques specifically for handling incomplete data, their current
state leaves much to be desired, both theoretically and practically.
Even evaluating SQL queries over incomplete databases -- a problem one
would expect to be solved after 40+ years of relational technology --
one gets results that make people say {\sl ``you can never trust the
  answers you get from [an incomplete] database''} \cite{date2005}. In
fact we know that SQL can produce every type of error imaginable when
nulls are present \cite{Ltods16}.

On the theory side, we appear to have a good understanding of what is
needed in order to produce correct results: computing {\em certain
  answers} to queries. These are answers that are true in all
complete databases that are compatible with the given incomplete database. This idea, that
dates back to the late 1970s as
well, has become {\em the} way of providing query answers in all
applications, from classical databases with incomplete information
\cite{IL84} to new applications such as data integration and exchange
\cite{Lenzerini02,ABLM14}, consistent query answering
\cite{bertossi-book}, ontology-based data access
\cite{CGLLR07}, and others. The reason these ideas have found limited
application in mainstream database systems is their
complexity. Typically, answering queries over incomplete databases
with certainty can be done efficiently for conjunctive queries or some
closely related classes, but beyond the complexity quickly grows to
intractable -- and sometimes even undecidable, see
\cite{pods14}. Since this cannot be tolerated by real life systems,
they resort to ad hoc solutions, which go for efficiency and sacrifice
correctness; thus bizarre and unexpected behavior occurs.

While even  basic problems related to
incompleteness in relational databases
remain unsolved, we now constantly deal with more varied types of
incomplete and inconsistent data. A prominent example is that of
probabilistic databases~\cite{probabilistic-book}, where the
confidence in a query answer is the total weight of the worlds that
support the answer. Just like certain answers, computing exact answer
probabilities is
usually intractable, and yet it has been the focus
of theoretical research.

The key challenge in addressing  the problem  of handling incomplete
and uncertain data is to provide
theoretical solutions that are {\em usable in practice}.  Instead of
proving more impossibility results, the field should urgently
address  what can actually be done efficiently.\\

Making theoretical results
applicable in practice is the biggest practical challenge for incomplete and
uncertain data. To move away from the focus on intractability and to
produce results of practical relevance, the PDM community needs to
address several challenges.

\begin{description}
\item[RDBMS technology in the presence of incomplete
      data.]
  It must be capable of finding query answers one can trust, and do so
  efficiently. But how do we find good quality query answers with
  correctness guarantees when we have theoretical intractability?  For
  this we need new approximation schemes, quite different from those
  that have traditionally been used in the database field.  Such
  schemes should provide guarantees that answers can be trusted, and
  should also be implementable using existing RDBMS technology.

  To make these scheme truly efficient, we need to address the issue
  of the performance of commercial RDBMS technology in the presence of
  incomplete data. Even query optimization in this case is hardly a
  solved problem; in fact commercial optimizers often do not perform
  well in the presence of nulls.

  \item[Models of uncertainty.]
  What is provided by current practical solutions is rather limited.
  Looking at relational databases, we know that they try to model
  everything with primitive null values, but this is clearly
  insufficient. We need to understand types of uncertainty that need
  to be modeled and introduce appropriate representation mechanisms.

  This, of course, will lead to a host of new challenges.  How do we
  store/represent richer kinds of uncertain information, that go well
  beyond nulls in RDBMSs?  Applications such as integration, exchange,
  ontology-based data access and others often need more (at the very
  least, marked nulls), and one can imagine many other possibilities
  (e.g., intervals for numerical values). This is closely related to
  the modelling data task described in Section~\ref{sec-mmd}.

  \item[Benchmarks for uncertain data.]
  What should we use as benchmarks when working with
  incomplete/uncertain data? Quite amazingly, this has not been
  addressed; in fact standard benchmarks tend to just ignore
  incomplete data, making it hard to test efficiency of solutions in
  practice.

  \item[Handling inconsistent data.] How do we make handling
  inconsistency (in particular, consistent query answering) work in
  practice? How do we use it in data cleaning? Again, there are many
  strong theoretical results here, but they concentrate primarily on
  tractability boundaries and various complexity dichotomies for
  subclasses of conjunctive queries, rather than practicality of query
  answering techniques. There are promising works on enriching
  theoretical \emph{repairs} with user preferences~\cite{StaworkoCM12},
  or ontologies~\cite{eiter-kr16}, along the lines of approaches
  described in Section \ref{sec-kedm}, but much more foundational work
  needs to be done before they can get to the level of practical
  tools.

  \item[Handling probabilistic data.] { The common models of
    probabilistic databases are arguably simpler and more restricted
    than the models studied by the Statistics and Machine Learning
    communities. Yet common complex models can be simulated by
    probabilistic databases if one can support expressive query
    languages~\cite{journals/pvldb/JhaS12}; hence, model complexity
    can be exchanged for query complexity. Therefore, it is of great
    importance to develop techniques for approximate query answering,
    on expressive query languages, over large volumes of data, with
    practical execution costs. {While the focus of the PDM community
      has been on deterministic and exact
      solutions~\cite{probabilistic-book}, we believe that more
      attention should be paid to statistical techniques with
      approximation guarantees such as the sampling approach typically
      used by the (Bayesian) Machine Learning and Statistics
      communities. In Section~\ref{sec-dmml} we further discuss the
      computational challenges of Machine Learning in the context of
      databases.}  }
\end{description}

\noindent
The theoretical challenges can
be split into three
groups.

\begin{description}
\item[Modeling.] We need to provide a solid theoretic
  basis for the practical modeling challenge above; this means
  understanding different types of uncertainty and their
  representations. As with any type of information stored in
  databases, there are lots of questions for the PDM community to work
  on, related to data structures, indexing techniques, and so on.

  There are other challenges related to modeling data. For instance,
  when can we say that some data is true? This issue is particularly
  relevant in crowdsourcing applications \cite{SarmaPW16,GrozMR15}: having data that looks
  complete does not yet mean it is true, as is often assumed.

  Yet another important issue addresses modeling query answers. How do
  we rank uncertain query answers? There is a tendency to divide
  everything into certain and non-certain answers, but this is often
  too coarse.

  The Programming Languages and Machine Learning communities have been
  investigating \emph{probabilistic
    programming}~\cite{DBLP:conf/popl/Goodman13} as a paradigm for
  allowing developers to easily program Machine Learning solutions.
  The Database community has been leading the development of paradigms
  for easy programming over large data volumes.  As discussed in
  detail later in Section~\ref{sec-dmml}, we believe that modern needs
  require the enhancement of the database technology with
  machine learning capabilities. In particular, an important challenge
  is to combine the two key capabilities (machine learning and data)
  via query languages for building statistical models, as already
  began by initial
  efforts~\cite{DBLP:conf/icdt/BaranyCKOV16,DBLP:conf/sigmod/CaiVPAHJ13}.

\item[Reasoning.]
  There is much work on this subject; see Section \ref{sec-kedm}
  concerning the need to develop next-generation reasoning tools for
  data management tasks. When it comes to using such tools with
  incomplete and uncertain data, the key challenges are: How do we do
  inference with incomplete data?  How do we integrate different types
  of uncertainty?  How do we learn queries on uncertain data?  What do
  query answers actually tell us if we run queries on data that is
  uncertain? That is, how results can be generalized from a concrete
  incomplete data set.

  \item[Algorithms.] To overcome high complexity, we often
  need to resort to approximate algorithms, but approximation
  techniques are different from the standard ones used in databases,
  as they do not just speed up evaluation but rather ensure
  correctness. The need for such approximations leads to a host of
  theoretical challenges. How do we devise such algorithms? How do we
  express correctness in relational data and beyond? How do we measure
  the quality of query answers? How do we take user preferences into
  account?
\end{description}

  While all the above are important research topics that need to be
addressed,
there are several that can be viewed as a
priority, not least because there is an immediate connection between
theory and practice. In particular, we need to pay close attention to the following issues: (1) understand what it means for answers to be right or wrong, and how to adjust the standard relational technology to ensure that wrong answers are never returned to the user; (2) provide, and justify, benchmarks for working with incomplete/uncertain data; (3) devise approximation algorithms for classes of queries known to be intractable; and (4) make an effort to achieve practicality of consistent query answering, and to apply it in data cleaning~scenarios.

\makeatletter{}\section{Knowledge-enriched Data Management}
\label{sec-kedm}

Over the past two decades we have witnessed a gradual shift from a world where
most data used by companies and organizations was regularly structured, neatly
organized in relational databases, and treated as complete, to a world where
data is heterogenous and distributed, and can no longer be treated as complete.
Moreover, not only do we have massive amounts of data; we also have very large
amounts of rich knowledge about the application domain of the data, in the form
of taxonomies or full-fledged ontologies, and rules about how the data should
be interpreted, among other things.
Techniques and tools for managing such complex information have been studied
extensively in Knowledge Representation, a subarea of Artificial Intelligence.  In
particular logic-based formalisms, such as description logics and different
rule-based languages, have been proposed and associated reasoning mechanisms
have been developed.  However, work in this area did not put a strong emphasis
on the traditional challenges of data management, namely huge volumes of data,
and the need to specify and perform complex operations on the data efficiently,
including both queries and updates.

Both practical and theoretical challenges arise when rich domain-specific
knowledge is combined with large amounts of data and the traditional data
management requirements, and the techniques and approaches coming from the
PDM community will provide important tools to address
them.  We discuss first the practical challenges.

\begin{description}
\item[Providing end users with flexible and integrated access to data.]
    A key requirement in dealing with complex, distributed, and heterogeneous
  data is to give end users the ability to directly manage such data.  This is
  a challenge since end users might have deep expertise about a specific domain
  of interest, but in general are not data management experts.  As a result,
  they are not familiar with traditional database techniques and technologies,
  such as the ability to formulate complex queries or update operations,
  possibly accessing multiple data sources over which the data might be
  distributed, and to understand performance implications.  Ontology-based data
  management has been proposed recently as a general paradigm to address this
  challenge.  It is based on the assumption that a domain ontology capturing
  complex knowledge can be used for data management by linking it to data
  sources using declarative mappings \cite{PLCD*08}.  Then, all information
  needs and data management requirements by end users are formulated in terms
  of such ontology, instead of the data sources, and are automatically
  translated into operations (queries and updates) over the data sources.  Open
  challenges are related to the need of dealing with distribution of data,
  of handling heterogeneity at both the intensional and extensional levels, of
  performing updates to the data sources via the ontology and the mappings, and
  in general of achieving good performance even in the presence of large
  ontologies, complex mappings, and huge amounts of data
  \cite{CGLLR07,KLTWZ11,GKKP*14,GoOP14,BiOr15}.

\item[Ensuring interoperability at the level of systems exchanging data.]
    Enriching data with knowledge is not only relevant for providing end-user
  access, but also enables direct inter-operation between systems, based on the
  exchange of data and knowledge at the system level.  A requirement is the
  definition of and agreement on standardized ontologies covering all necessary
  aspects of specific domains of interest, including multiple modalities such
  as time and space.  A specific area where this is starting to play an
  important role is e-commerce, where standard ontologies are already
  available \cite{Hepp15}.

\item[Personalized and context-aware data access and management.]
    Information is increasingly individualized and only fragments of the
  available data and knowledge might be relevant in specific situations or for
  specific users.  It is widely acknowledged
    that it is necessary to provide mechanisms on the one hand for characterizing
  contexts (as a function of time, location, involved users, etc.), and on the
  other hand for defining which fragments of data and/or knowledge should be
  made available to users, and how such data needs to be
  pre-processed/filtered/modified, depending on the actual context and the
  knowledge available in that context.  The problem is further complicated by
  the fact that both data and knowledge, and also contextual information, might
  be highly dynamic, changing while a system evolves \cite{BaKP12,KlGu10}.
    Heterogeneity needs to be dealt with, both with respect to the modeling formalism and with respect
  to the modeling structures chosen to capture a specific real-world
  phenomenon.

\item[Bringing knowledge to data analytics and data extraction.]
    Increasing amounts of data are being collected to perform complex analysis
  and predictions.  Currently, such operations are mostly based on data in
  ``raw'' form, but there is a huge potential for increasing their
  effectiveness by enriching and complementing such data with domain knowledge,
  and leveraging this knowledge during the data analytics and extraction
  process. Challenges include choosing the proper formalisms for expressing
  knowledge about both raw and aggregated/derived data, developing
  knowledge-aware algorithms for data extraction and analytics, in particular
  for overcoming low data quality, and dealing with exceptions and outliers.
  
\item[Making the management user friendly.]
    Systems combining large amounts of data with complex knowledge are themselves
  very complex, and thus difficult to design and maintain.  Appropriate tools
  that support all phases of the life-cycle of such systems need to be designed
  and developed, based on novel user interfaces for the various
  components. Such tools should themselves rely on the domain knowledge and the
  sophisticated inference services over such knowledge to improve user
  interaction, in particular for domain experts as opposed to IT or data
  management experts.  Supported tasks should include design and
  maintenance of ontologies and mappings (including debugging support), query
  formulation, explanation of inference, and data and knowledge exploration
  \cite{FGTT11,LMRST15,DrLB15,DBLP:journals/ws/ArenasGKMZ16}.
\end{description}

To provide adequate solutions to the above practical challenges, several key
theoretical challenges need to be addressed, requiring a blend of formal
techniques and tools traditionally studied in data management, with those
typically adopted in knowledge representation in AI.

\begin{description}
\item[Development of reasoning-tuned DB systems.]  Such systems will require
  new/improved database engines optimized for reasoning over large amounts of
  data and knowledge, able to compute both crisp and approximate answers, and
  to perform distributed reasoning and query evaluation.  To tune such systems
  towards acceptable performance, new cost models need to be defined, and new
  optimizations based on such cost models need to be developed.
\item[Choosing/designing the right languages.]  The languages and formalisms
  adopted in the various components of knowledge-enriched data management
  systems
            have to support different types of knowledge and data, e.g., mixing open and
  closed world assumption, and allowing for representing temporal, spatial, and
  other modalities of information \cite{CaDL08,BLMS11,BCLW14,AKRZ14,NgOS16}.
  It is well understood that the requirements in terms of expressive power for
  such languages would lead to formalisms that make the various inference tasks
  either undecidable or highly intractable.  Therefore, the choice or design of
  the right languages have to be pragmatically guided by user and application
  needs.
\item[New measures of complexity.]  To appropriately assess the performance of
  such systems and be able to distinguish easy cases that seem to work well in
  practice from difficult ones, alternative complexity measures are required
  that go beyond the traditional worst-case complexity.  These might include
  suitable forms of average case or parameterized complexity, complexity taking
  into account data distribution (on the Web), and forms of smoothed analysis.
\item[Next-generation reasoning services.]  The kinds of reasoning services
  that become necessary in the context of knowledge-enriched data management
  applications go well beyond traditional reasoning studied in knowledge
  representation, which typically consists of consistency checking,
  classification, and retrieval of class instances.  The forms of reasoning
  that are required include processing of complex forms of queries in the
  presence of knowledge, explanation (which can be considered as a
  generalization of provenance), abductive reasoning, hypothetical reasoning,
  inconsistency-tolerant reasoning, and defeasible reasoning to deal with
  exceptions.  Forms of reasoning with uncertain data, such as probabilistic or
  fuzzy data and knowledge will be of particular relevance, as well as
  meta-level reasoning.  Further, it will be necessary to develop novel forms
  of reasoning that are able to take into account non-functional requirements,
  notably various measures for the quality of data (completeness, reliability,
  consistency), and techniques for improving data quality.  While such forms of
  reasoning have already begun to be explored individually (see, e.g.,
  \cite{ElKS06,BoDP15}, much work remains to bring them together, to
  incorporate them into data-management systems, and to achieve the necessary
  level of performance.
\item[Incorporating temporal and dynamic aspects.]  A key challenge is
  represented by the fact that data and knowledge is not static, and changes
  over time, e.g., due to updates on the data while taking into account
  knowledge, forms of streaming data, and more in general data manipulated by
  processes.  Dealing with dynamicity and providing forms of inference (e.g.,
  formal verification) in the presence of both data and knowledge is extremely
  challenging and will require the development of novel techniques and tools
  \cite{Calvanese-et-al-PODS-2013,AKRZ14}.
\end{description}

In summary, incorporating domain-specific knowledge to data management is both
a great opportunity and a major challenge.  It opens up huge possibilities for
making data-centric systems more intelligent, flexible, and reliable, but
entails computational and technical challenges that need to be overcome.
We believe that much can be achieved in the coming years. 
Indeed, the increasing interaction of the PDM and the Knowledge
Representation communities has been very fruitful, particularly by attempting
to understand the similarities and differences between the formalisms and
techniques used in both areas, and obtaining new results building on mutual
insights.  Further bridging this gap by the close collaboration of both areas
appears as the most promising way of fulfilling the promises of
Knowledge-enriched Data Management.

\makeatletter{}\section{Data Management and Machine Learning}\label{sec-dmml}

We believe that research that combines Data Management (DM) and
Machine Learning (ML) is especially important, because these fields
can mutually benefit from each other.  Nowadays, systems that emerge
from the ML community are strong in their capabilities of statistical
reasoning, and systems that emerge from the DM community are strong in
their support for data semantics, maintenance and scale. This
complementarity in assets is accompanied by a difference in the core
mechanisms: the PDM community has largely adopted methodologies driven
by logic theory, while the ML community centralized around probability
theory and statistics.  Yet, modern applications require systems that
are strong in \emph{both} aspects, providing a thorough and sophisticated
management of data while incorporating its inherent statistical
nature.  We envision a plethora of research opportunities in the
intersection of PDM and ML. We outline several directions, which we
classify into two categories: \emph{DM for ML} and \emph{ML~for~DM}.\\

The category \emph{DM for ML} includes directions that are aimed at the
enhancement of ML capabilities by exploiting properties of the data.
Key challenges are as follows.

\def\itemttl#1{\e{#1.}\,\,}

\begin{description}
\item[Feature Generation and Engineering.]  Feature engineering refers
  to the challenge of designing and extracting signals to provide to
  the general-purpose ML algorithm at hand, in order to properly
  perform the desired operation (e.g., classification or
  regression). This is a critical and time-consuming
  task~\cite{DBLP:dblp_conf/sigmod/ZhangKR14,DBLP:journals/tvcg/KandelPHH12},
  and a central theme of modern ML methodologies, such as kernel-based
  ML, where complex features are produced implicitly via kernel
  functions \cite{scho_lw}, and deep learning, where low-level
  features are combined into higher-level features in a hierarchical
  manner \cite{bengio13}.  Unlike general-purpose ML
    algorithms that view features as numerical values, the database
    has access to, and understanding of, the \emph{queries} that
    transform raw data into these features.
            Thus, PDM can contribute to feature engineering in
  various ways, especially on a semantic level, and provide solutions
  to problems such as the following: How to develop effective
  languages for query-based feature creation? How to use such
  languages for designing a set of complementary, non-redundant
  features optimally suited for the ML task at hand?  
    Is a given language suitable for a
  certain class of ML tasks? Important criteria
    for the goodness of a feature language include the risks of
    \emph{underfitting} and \emph{overfitting} the training data, as well
    as the computational complexity of evaluation (on both training
    and
    test data). The PDM community has already
  studied problems of a similar
  nature~\cite{cohen_et_al:LIPIcs:2015:4981,DBLP:journals/tods/KimelfeldK14,DBLP:journals/jacm/GottlobS10}.

The premise of deep (neural network) learning is that the model has
sufficient expressive power to work with only \emph{raw, low-level
  features}, and to realize the process of high-level feature
generation in an automated, data-driven manner~\cite{bengio13}. This
brings a substantial hope for reducing the effort in manual feature
engineering. 
    Is
there a general way of solving ML tasks by applying deep
  learning directly to the database (as has already been done, for
  example, with \emph{semantic
    hashing}~\cite{salakhutdinov09})? Can database queries (of
  different languages) complement neural networks by means of
  expressiveness and/or efficiency? And if so, where lies
    the boundary between the level of feature engineering and the
    complexity of
    the network?

\item[Large-Scale Machine Learning.]  Machine learning is nowadays
  applied to massive data sets of considerable size, including
  potentially unbounded streams of data \cite{domingos00}. Under such
  conditions, an effective data management and the use of appropriate
  data structures that offer the learning algorithm fast access to the
  data are major prerequisites for realizing model induction (at
  training time) and inference (at prediction time) in a
  time-efficient and space-efficient manner \cite{prabhu14}. Research
  along this direction has amplified in recent years and includes, for
  example, the use of hashing \cite{weinberger09}, Bloom filters
  \cite{cisse13}, and tree-based data structures
  \cite{dembczynski10,choromanska15} in learning algorithms. As
  another example, lossless compression of large datasets, as featured
  by \emph{factorized databases}~\cite{DBLP:journals/tods/OlteanuZ15},
  have been shown to dramatically reduce the execution cost of
  machine-learning tasks.  Also related is work on distributed machine
  learning, where data storage and computation is accomplished in a
  network of distributed units \cite{agarwal14}, and the support of
  machine learning by data stream management systems \cite{morales15}.

\item[Complexity Analysis.]  Over the years, the PDM community has
  established a strong machinery and repertoire for fine-grained
  analysis of querying complexity.  Examples include different notions
  of sensitivity to queries such as data
  complexity~\cite{DBLP:conf/stoc/Vardi82}, 
                parameterized
  complexity~\cite{DBLP:journals/jcss/PapadimitriouY99},
  dichotomies/trichotomies in
  complexity~\cite{DBLP:journals/jacm/DalviS12,DBLP:conf/pods/Kimelfeld12,DBLP:conf/pods/KoutrisW15},
  and sensitivity to data properties such as
  acyclicity~\cite{DBLP:conf/icalp/AmarilliBS15}.  Complexity analysis
  of such granularity is highly desirable for the ML community,
  especially for analyzing learning algorithms that involve various
  parameters like input dimension (number of features), output
  dimension, and number of training examples~\cite{jasinska16}.
  Ideally, such analyses give rise to novel techniques and data
  structures for important special cases.  Results along this
  direction, connecting DM querying complexity and ML training
  complexity, have been recently
  shown~\cite{DBLP:conf/sigmod/SchleichOC16}.
\end{description}

The motivation for the directions in the second category, \emph{ML for
DM}, is that of
strengthening core data-management capabilities with
ML. Traditionally, data management systems have supported a core set
of querying operators (e.g., relational algebra, grouping and
aggregate functions, recursion) that are considered as the common
requirement of applications. We believe that this core set should be
revisited, and specifically that it should be extended with common ML
operators. 

As a prominent example, motivated by the proliferation of available
and valuable textual resources, various formalisms have been proposed
for incorporating text extraction in a relational
model~\cite{fagin2015document,FaginKRV14,DBLP:conf/vldb/ShenDNR07}.
However, unlike structured data, textual resources are associated with
a high level of uncertainty due to the uncontrolled nature of the
content and the imprecise nature of natural language processing. Therefore, ML
techniques are required to distill reliable information
from text. 

We believe that incorporating ML is a natural evolution for PDM.
Database systems that incorporate statistics and ML have
  already been
  developed~\cite{DBLP:conf/vlds/NiuZRS12,DBLP:journals/pvldb/ShinWWSZR15,DBLP:conf/sigmod/ArefCGKOPVW15}.
Query languages have traditionally been designed with emphasis on
being \emph{declarative}: a query states how the answer should logically
relate to the database, not how it is to be computed algorithmically.
Incorporating ML introduces a higher level of declarativity, where one
states how the end result should behave (via examples), but not
necessarily which query is deployed for the task. In that spirit, we
propose the following directions for relevant PDM research.

\begin{description}
\item[Unified Models.] An important role of the PDM community is
in establishing common formalisms and semantics for the database
community. It is therefore an important opportunity to establish the
``relational algebra'' of data management systems with built-in
ML/statistics operators.

\item[Lossy Optimization.] From the early days of Selinger's query
  planning~\cite{DBLP:conf/sigmod/SelingerACLP79} and Yannakakis's
  acyclic-join algorithm~\cite{DBLP:conf/vldb/Yannakakis81}, the focus
  of the PDM community has been on \emph{lossless} optimization, that is,
  optimization that leaves the end result intact. As
    mentioned in
  Section~\ref{sec-mds}, in some scenarios it makes sense to apply
  \emph{lossy} optimization that guarantees only an approximation of the
  true answer. Incorporating ML into the query model gives further
opportunities for lossy optimization, as training
  paradigms are typically associated with built-in quality (or
  ``risk'') functions. Hence, we may consider reducing the execution
  cost if it entails 
  a bounded impact on the quality of the end result~\cite{akdere11}.
  For example, Riondato et al.~\cite{riondato11} develop a method for
  random sampling of a database for estimating the selectivity of a
  query. Given a class of queries, the execution of any query in that
  class on the sample provides an accurate estimate for the
  selectivity of the query on the original large database.

\item[Confidence Estimation.] Once statistical and ML components
are incorporated in a data management system, it becomes crucial to
properly estimate the \emph{confidence} in query
answers~\cite{DBLP:journals/pvldb/ShinWWSZR15}, as such a confidence
offers a principled way of controlling the balance between precision
and recall. It is then an important direction to establish
probabilistic models that capture the combined process and allow to
estimate probabilities of end results.  For example, by applying the
notion of the Vapnik-Chervonenkis dimension, an important theoretical
concept in generalization theory, to database queries, Riondato et
al.\ \cite{riondato11} provide accurate bounds for their selectivity
estimates that hold with high probability; moreover, they show the
error probability to hold simultaneously for the selectivity estimates
of all queries in the query class.  In general, this direction can
leverage the past decade of research on probabilistic
databases~\cite{DBLP:journals/jacm/DalviS12,DBLP:journals/jacm/KimelfeldR14,DBLP:conf/icdt/BaranyCKOV16,DBLP:series/sfsc/KimelfeldS13},
which can be combined with theoretical frameworks of machine learning,
such as PAC (Probably Approximately Correct) learning
\cite{valiant84}.
\end{description}

\noindent
Altogether, we have a plethora of research problems,  on improving machine learning with data management techniques (DM
for ML), and on strengthening data management technologies with capabilities
of machine learning (ML for DM).  The required methodologies and
formal foundations span a variety of related fields such as logic,
formal languages, computational complexity, statistical analysis, and
distributed computing.  We phrased the
directions as theoretically oriented; but obviously, each of them is
coming with the practical challenge of devising effective solutions
over real systems, and on real-life datasets and benchmarks.

\makeatletter{}\section{Process and Data}\label{sec-pd}

Many forms of data evolve over time, and most processes access and
modify data sets.  Industry works with massive volumes of evolving
data, primarily in the form of transactional systems and 
Business Process Management (BPM) 
systems.  Research into 
basic questions about systems that combine process and data
has been growing over the past decade,
including
the development of several formal models,
frameworks for comparing their expressive power,
approaches to support verification of behavioral properties,
and query languages for process schemas and instances.

Over the past half century, computer science research has studied
foundational issues of process and of data mainly as separated
phenomena.  

In recent years, data and process have been studied together in two
significant areas: scientific workflows and
data-aware BPM \cite{Hull-Su-NSF-report-2009}.  
Scientific workflows focus on
enabling repeatability and reliability of processing flows involving
large sets of scientific data.
In the 1990's and 00's, 
foundational research in this area helped to establish the basic
frameworks for supporting these workflows, to enable the systematic
recording and use of provenance information, and to support systems
for exploration that involve multiple runs of a workflow with
varying configurations \cite{davidson-freire:sci-wf:sigmod-2008}.  The work on
scientific workflows can also play a role in enabling process support
for big data analytics, especially as industry begins to create
analytics flows that can be repeated, with relatively minor variation,
across multiple applications and clients.

Foundational work on data-aware BPM was launched in
the mid-00's \cite{Bhattacharya-et-al-BPM-2007, DHPV-analysis-et-al:ICDT-09}, 
enabled in part by IBM's ``Business Artifacts''
model for business process \cite{Nigam-Caswell-2003}, that combines data and process in a holistic
manner.
Deutch and Milo \cite{Milo:PODS-2011} provide a survey and comparison
of several of the most important early models and results on
process and data.
One variant of the business artifact model,
which is formally defined around logic rather than Petri-nets,
has provided the conceptual basis
for the recent OMG Case Management Model and Notation standard
\cite{GSM-and-CMMN:2012}.
Importantly, the artifact-based perspective has formed the basis for a vibrant
body of work centered around verification of systems that support
processes involving large-scale data \cite{Calvanese-et-al-PODS-2013,DeutschHV14}.
The artifact-based perspective is also beginning to enable a more unified
management of the interaction of business processes 
and legacy data systems \cite{Sun-Su-Yang-TMIS-2016}.
Importantly, there is strong overlap between the artifact-based approach and
core building blocks of the ``shared ledger'' approach to supporting business
(and individual) interactions around the exchange of goods and services, 
as embodied initially by the Blockchain paradigm of Bitcoin 
\cite{Tschorsch-Scheuermann-Blockchain}.

Foundational work in the area of process and data has the potential for 
continued and expanded
impact in the following six practical challenge areas.

\begin{description}
\item[Automating manual processes.]
Most business processes still rely on substantial manual effort. 
In the case of ``back-office'' processing, Enterprise Resource Planning
systems such as SAP automatically perform the bulk of the work,
e.g., for applications in finance and human resource management.
But there are still surprisingly many ``ancillary processes''
that are performed manually, e.g., to process new bank accounts or
newly hired employees.
In contrast, business processes that involve substantial human
judgement, such as complex sales activities or the transition of IT services
from one provider to another, 
are handled today in largely {\em ad hoc}
and manual ways, with spreadsheets as the workflow management tool of choice.

\item[Evolution and migration of Business Processes.]
Managing change of business processes remains 
largely manual,
highly expensive, time consuming, and risk-prone.
This includes
deployment of new business process platforms, evolution of business processes,
and integration of business processes after mergers.

\item[Business Process compliance and correctness.]
Compliance with government regulations and corporate policies is a rapidly growing challenge,
e.g., as governments attempt to enforce policies around
financial stability and data privacy.
Ensuring compliance is largely manual today, and involves
understanding how regulations can impact or define
portions of business processes,
and then verifying that process executions will comply.

\item[Business Process interaction and interoperation.]
Managing business processes that flow across enterprise boundaries
has become increasingly important with globalization of business
and the splintering of business activities across numerous companies.
While routine services such as banking money transfer are largely
automated, most interactions between businesses are less standardized
and require substantial manual effort to set up, maintain, and troubleshoot.
The recent industrial interest in shared ledger
technologies highlights the importance of this 
area and provides new motivation for developing foundational results
for data-aware processes.

\item[Business Process discovery and understanding.]
The field of Business Intelligence, which provides techniques
for mining and analyzing information about business operations,
is essential to business success.  Today this field
is based on a broad variety of largely {\em ad hoc} and manual 
techniques 
\cite{dayal-ETL-EDBT-invited-2009},
with associated costs and potential for error.
One important direction on understanding processes focuses on
viewing process schemas and process instances as data,
and enabling declarative query languages against them
\cite{Milo-et-al:querying-process-instances}. 
More broadly, techniques from 
Multi-model Data Management (Section \ref{sec-mmd}),
Data Management and Machine Learning (Section \ref{sec-dmml}),
and Uncertain Data (Section \ref{sec-ui})
are all relevant here because of
(respectively)
the heterogeneity of data about and produced by processes,
the importance of anticipating undesirable outcomes and mitigating,
and the fact that the information stored about processes is often incomplete.

\item[Workflow and Business Process usability.]
The operations of 
medium- and large-sized enterprises are highly complex,
a situation enabled in part by the power of computers to
manage huge volumes of data, transactions, and processing all at tremendously
high speeds. 
This raises questions relating to 
Managing Data at Scale (Section \ref{sec-mds}).
Furthermore, enabling humans to understand and work effectively 
to manage large numbers of 
processes remains elusive,
especially when considering the interactions between process, data (both
newly created and legacy), resources, the workforce, and business partners.
\end{description}

The above practical BPM challenges raise key research challenges 
that need to be addressed using approaches that include
mathematical and algorithmic frameworks and tools.

\begin{description}
\item[Verification and Static Analysis.]
Because of the infinite state space inherent in data-aware processes
\cite{Calvanese-et-al-PODS-2013,DeutschHV14},
verification currently relies on faithful abstractions reducing the problem to
classical finite-state model checking. However, the
work to date can only handle restricted classes of applications,
and research is needed to develop more powerful abstractions enabling a variety of
static analysis tasks for realistic data-aware processes. 
Incremental verification techniques are needed, as well as techniques that
enable modular styles of verification that support ``plug and play''
approaches.
This research will be relevant to the first four practical challenges.

\item[Tools for Design and Synthesis.]
Formal languages (e.g., context-free) had a profound impact on compiler theory and
programming languages. Dependency theory and normal forms had a profound
impact on relational database design.
But 
there is still no robust framework that supports 
principled design of business processes 
in the larger context of data, resources, and workforce.
Primitive operators for creating and modifying data-aware process schemas
will be an important starting point;
the ultimate goal is partial or full synthesis of process from
requirements, goals, and/or regulations.
This research will be relevant to the first, second, fourth, and sixth
practical challenges.

\item[Models and semantics for views, interaction, and interoperation.]
The robust understanding of database views has enabled
advances in simplification of data access, data sharing, exchange, integration, and privacy, as well as query optimization.
A robust theory of views for data-aware business processes
has similar potential.
For example, it could support
a next generation of data-aware service composition techniques
that includes practical verification capabilities.
Frameworks that enable comparison of process models
(e.g., \cite{ABV:workflow-comparisons}) can provide an important
starting point for this research.
This research will be relevant to all of the practical challenges.

\item[Analytics for Business Processes.]
The new, more holistic perspective of data-aware processes 
can help to provide a new foundation for the field of business intelligence.
This can include new approaches for instrumenting processes to simplify
data discovery \cite{Liu-Vaculin:artifact-centric-monitoring}, 
and new styles of modularity and hierarchy in both
the processes and the analytics on them.
\end{description}

Research in process and data
will require
on-going extensions of the traditional approaches, 
on both the database and process-centric sides.
New approaches may include models for the creation and maintenance of
interoperations between (enterprise-run) services;
semi-structured and unstructured forms of data-aware business process 
(cf.\ noSQL);
new abstractions to enable verification over infinite-state systems;
and new ways to apply machine learning.
More broadly, a new foundational model for modern BPM 
may emerge, which builds on the artifact and shared-ledger approaches
but facilitates a multi-perspective understanding, analogous to the way
relational algebra and calculus provide two perspectives on data querying.

One cautionary note is that research in the area of process and data today
is hampered by a lack of large sets of examples, e.g., 
sets of process schemas that include explicit specifications concerning
data, and process histories that include how data sets were used and affected.
More broadly, increased collaboration between
PDM researchers, applied BPM researchers,  and 
businesses would enable more rapid progress towards
resolving the  concrete problems in BPM faced by industry today.

\makeatletter{}\section{Human-related data and ethics}\label{sec-ethics}

More and more ``human-related'' data is massively generated, in
particular on the Web and in phone apps. Massive data analysis, using
data parallelism and machine learning techniques, is applied to this
data to generate more data. We, individually and collectively, are
losing control over this data. We do not know the answers to questions
as important as: Is my medical data really available so that I get
proper treatment? Is it properly protected?  Can a private company
like Google or Facebook influence the outcome of national elections?
Should I trust the statistics I find on the Web about the crime rate
in my neighborhood?

Although we keep eagerly consuming and enjoying more new Web services
and phone apps, we have growing concerns about criminal behavior on
the Web, including racist, terrorist, and pedophile sites; identity theft;
cyber-bullying; and cyber crime. We are also feeling growing
resentment against intrusive government practices such as massive
e-surveillance even in democratic countries, and against aggressive
company behaviors such as invasive marketing, unexpected
personalization, and cryptic or discriminatory business decisions.

Societal impact of big data technologies is receiving significant
attention in the popular press~\cite{compas}, and is under active
investigation by policy makers~\cite{WHMay2016} and
legal scholars~\cite{BarocasSelbst,Kroll2017,lerman2013big}.  It is
broadly recognized that this technology has the potential to improve
people's lives, accelerate scientific discovery and innovation, and
bring about positive societal change.  It is also clear that the same
technology can in effect limit business faithfulness to legal and
ethical norms, and that it raises the danger of ``algocracy'' --- rule
by incontestable algorithms~\cite{Danaher}.  And while many of the
issues are political and economical, technology solutions must play an
important role in enabling our society to reap ever-greater benefits
from big data, while keeping it safe from the risks.

Since the 20th century, an important societal challenge for computer
science has been the efficient and effective management of larger and
larger volumes of data generated by computer applications. The data
management R\&D was therefore primarily driven by the study of data
models and by system performance, and has been immensely successful.
We have developed technology needed to manage huge volumes of data.
We believe that the main inspiration for the data management field in
the 21st century comes from the management of human-related data, with
an emphasis on solutions that satisfy ethical requirements.

In the remainder of this section, we will present several facets of
ethical data management. Some of these have been receiving attention
of specific research communities within computer science, most notably
data mining and machine
learning~\cite{DBLP:conf/kdd/HajianBC16,DBLP:conf/icml/ZemelWSPD13},
privacy~\cite{D11}, and systems and internet
measurement~\cite{DBLP:journals/popets/DattaTD15,DattaSZ16,DBLP:conf/www/EnglehardtREZMN15,Letal14}.
The data management community (both theory and systems) could greatly
contribute to ethical data management.  There are significant
opportunities specifically for PDM, which
can help to clarify issues,
to develop precise models of key characteristics related to ethical
data management,
and to understand the inherent feasibility of proposed
approaches and algorithms.

\begin{description}
\item[Responsible data analysis.]
  Human-related data analysis needs to be ``responsible'' --- to be
  guided by humanistic considerations and not simply by performance or
  by the quest for profit. The notion of responsible data analysis is
  considered generally in~\cite{ASM16,DBLP:conf/edbt/StoyanovichAM16}
  and was the subject of a recent Dagstuhl
  seminar~\cite{dagstuhl/16291}.  We now outline several important
  aspects of the problem, especially those where we see opportunities
  for involvement by PDM.  This list is by no means exhaustive, and
  does not include, e.g., privacy~\cite{D11}, which is already
  receiving significant attention of the PDM community.

  \begin{itemize}
\item {\bf Fairness.}  Responsible data analysis requires that both
    the raw data and the computation be ``fair'', i.e. not biased
    \cite{Detal12,DBLP:conf/kdd/HajianBC16,DBLP:conf/icml/ZemelWSPD13}.
    There are technical challenges in specifying fairness.  For
    example, there are different statistical fairness criteria, such
    as individual vs.\ group, and a variety of specific
    formulations~\cite{DBLP:journals/corr/Zliobaite15a}. Some of this
    work draws on deep connections with differential
    privacy~\cite{Detal12}.  There is currently no consensus as to
    which classes of fairness measures, and which specific
    formulations, are appropriate for various data analysis
    tasks~\cite{DBLP:journals/corr/TramerAGHHHJL15}.  Work is needed
    to formalize the measures and understand the relationships between
    them.  An important research direction for the PDM and the
    database systems communities is to understand how fairness
    quantification and enforcement can be pushed closer to the data,
    and interleaved with data manipulation operations, in relational
    algebra and beyond.

  \item {\bf Transparency and accountability.} Responsible data
    analysis practices must be
    transparent~\cite{DBLP:journals/popets/DattaTD15,DattaSZ16,DBLP:journals/cacm/Sweeney13},
    allowing a variety of stakeholders, such as end-users, commercial
    competitors, policy makers, and the public, to scrutinize the data
    collection and analysis processes, and to interpret the outcomes.
    Transparency is valuable in its own right, but also enables
    accountability, i.e., checking that a system behaves in accordance
    with legal norms, acceptable business practices, and its own
    stated commitments~\cite{Kroll2017}.  While transparency is
    clearly related to open source and open data, complete disclosure
    of data and code may be impossible, e.g., for privacy or business
    considerations.  Interesting research challenges that can be
    tackled by PDM include using provenance to shed light on data
    collection and analysis practices, supporting semantic
    interrogation of data analysis methods and pipelines, and
    providing explanations in various contexts, including
    knowledge-based systems and deep learning.

  \item {\bf Diversity.}  Big data technology poses significant risks
    to those it overlooks~\cite{lerman2013big}.  Diversity
    \cite{A09,DBLP:journals/pvldb/DrosouP12,DBLP:journals/pvldb/Wu0HZL15}
    requires that not all attention be devoted to a limited set of
    objects, actors or needs.  In an on-line dating platform, a
    crowdsourcing marketplace or a funding platform, it is often the
    case that a small subset of the items dominates rankings and so is
    given unfair advantage.
                Indeed, in crowdsourcing, diversity of opinion is common \cite{SarmaPW16,GrozMR15}
    and one of the four elements required to form a wise
    crowd~\cite{DBLP:journals/pvldb/Wu0HZL15}.  Diversity is related
    to serendipity: a search engine should not exclude from results
    interesting pages from an entire area, simply because they are not
    popular enough, or do not match the core interests of the user.
    Despite its importance, diversity has been studied in a limited
    set of scenarios, and rarely when data is about people.  The PDM
    community can contribute, for instance, to understanding the
    connections between diversity and fairness, and to develop methods
    to manage trade-offs between diversity and conventional measures
    of accuracy.

  \end{itemize}
\item[Verifying data responsibility.]  A
  grand challenge for the community is to develop verification
  technology to enable a new era of responsible data.  One can first
  envision research towards developing tools to help users understand
  data analysis results (e.g., on the Web), and to verify them.
  Verification of massive data analysis might involve code
  verification and/or systematic testing.  Code verification has been
  studied a lot to check for security (see e.g.\ \cite{CK14}), safety,
  and privacy, but rarely for verifying responsibility properties. The
  testing of statistical properties is now an old field, that needs to
  progress to allow testing on the huge data volumes found on the Web.

  One can also envision tools that help analysts, who are typically
  not computer scientists nor experts in statistics, to realize
  responsible data analysis ``by design''. Such a tool should
  accompany the analysts starting from the selection of data, raising
  issues such as the fairness of that selection, and of the processing
  that is performed.
\item[Data quality and access control on
    the Web.]  The evaluation of data quality on the Web is an issue
  of paramount importance when our lives are increasingly guided and
  determined by data found on the Web.  We would like to know whether
  we can trust some particular data we found. Has it been correlated
  to other data?  Is it controversial?  We would like to be able to
  evaluate the quality of information and the trustworthiness of
  sites.  This aspect leads to reasoning about human-related data,
  which can of course benefit from research on Uncertain Information
  (Section \ref{sec-ui}). The human-related nature of data brings new
  dimensions; this is not only about truth but also about opinions,
  bias, sentiments, etc.

  Once some data has been published on the Web, there is currently no
  built-in means of specifying where it comes from, who should be
  allowed to read it, update it, or for what purposes it can be used.
  Privacy has been already intensively studied, see e.g.
  \cite{D11,FBetal10}, but mostly in a centralized context.  Research
  is needed towards supporting access control on the Web. It may build
  for instance on cryptography, blockchain technology, or distributed
  access control \cite{Metal15}.
\item[Personal information management
    system.]  A Personal Information Management System (PIMS) is a (cloud)
  system that manages all the information of a person. By returning
  part of the data control to the person, these systems tend to better
  protect privacy, re-balance the relationship between a person and
  the major internet companies in favor of the person, and in general
  facilitate the protection of ethical values \cite{AAK15}. By making
  the person the focus, the PIMS approach tends to bring new
  challenges to existing topics in data management, e.g.,
  Knowledge-enriched Data Management (Section \ref{sec-kedm}) and
  Uncertain Information (Section \ref{sec-ui}).
\end{description}

Ethical data management raises new issues for
  computer science in general and for data management in
  particular. Because the data of interest is typically human-related,
  the research also includes aspects from other sciences, notably,
  cognitive science, psychology, neuroscience, linguistics, sociology,
  and political sciences. The ethics component also leads to
  philosophical considerations. In this setting, researchers have a
  chance for major societal impact, and so they need to interact with
  policy makers and regulators, as well as with the media and user
  organizations.
 
\makeatletter{}\section{Looking Forward}

As illustrated in the preceding sections,
the principled, mathematically-based approach to the
study of data management problems is providing 
conceptual foundations, deep insights, and much-needed clarity.
This report describes a representative, but by no means exhaustive,
family of areas where research on the
Principles of Data Management (PDM)
can help to shape
our overall approach to working with data as it arises
across an increasingly broad array of application areas.

The Dagstuhl workshop highlighted two important trends that have been
accelerating in the PDM
community over the past several years.  
The
first is the increasing embrace of neighboring disciplines, including
especially
Machine Learning, Statistics, Probability, and Verification, both to
help resolve new challenges, and to bring new perspectives to them.
The second is the increased focus on obtaining positive results, that
enable the use of mathematically-based insights in practical settings.
We expect and encourage these trends to continue in the coming years.

The need for precise and robust approaches 
for increasingly varied forms of data management 
continues to intensify, given the
fundamental and transformational role of
data in our modern society,
and given the continued expansion of
technical, conceptual, and ethical
data management challenges.
There is an associated and on-going expansion in
the family of
approaches and techniques
that will be relevant to PDM research.
The centrality of data management across numerous 
application areas is an opportunity both
for PDM researchers to 
embrace techniques and perspectives from adjoining research areas,
and for researchers from other areas to
incorporate techniques and perspectives from PDM.
Indeed, we hope that this report can substantially strengthen
cross-disciplinary research between the PDM and
neighboring theoretical communities and, moreover,
the applied and systems research communities 
across the many application areas that rely on data in one form or another.

\bibliographystyle{abbrv}
\bibliography{whitepaper}

\begin{thebibliography}{100}

\bibitem{DBLP:journals/cacm/AbadiAABBCCDDFG16}
D.~Abadi, R.~Agrawal, A.~Ailamaki, M.~Balazinska, P.~A. Bernstein, M.~J. Carey,
  S.~Chaudhuri, J.~Dean, A.~Doan, M.~J. Franklin, J.~Gehrke, L.~M. Haas, A.~Y.
  Halevy, J.~M. Hellerstein, Y.~E. Ioannidis, H.~V. Jagadish, D.~Kossmann,
  S.~Madden, S.~Mehrotra, T.~Milo, J.~F. Naughton, R.~Ramakrishnan, V.~Markl,
  C.~Olston, B.~C. Ooi, C.~R{\'{e}}, D.~Suciu, M.~Stonebraker, T.~Walter, and
  J.~Widom.
\newblock The {B}eckman report on database research.
\newblock {\em Commun. {ACM}}, 59(2):92--99, 2016.

\bibitem{AAK15}
S.~Abiteboul, B.~Andr{\'{e}}, and D.~Kaplan.
\newblock Managing your digital life.
\newblock {\em Commun. {ACM}}, 58(5):32--35, 2015.

\bibitem{ABV:workflow-comparisons}
S.~Abiteboul, P.~Bourhis, and V.~Vianu.
\newblock Comparing workflow specification languages: {A} matter of views.
\newblock {\em {ACM} Trans. Database Syst.}, 37(2):10, 2012.

\bibitem{dagstuhl/16291}
S.~Abiteboul, G.~Miklau, J.~Stoyanovich, and G.~Weikum, editors.
\newblock {\em Data, Responsibly}, volume 16291 of {\em Dagstuhl Seminar
  Proceedings}. Schloss Dagstuhl -- LZI, 2016, forthcoming.

\bibitem{ASM16}
S.~Abiteboul and J.~Stoyanovich.
\newblock Data, responsibly.
\newblock ACM SIGMOD Blog, 20 November 2015.
\newblock \url{https://hal.inria.fr/hal-01248054/}.

\bibitem{afrati11:_optim}
F.~N. Afrati and J.~D. Ullman.
\newblock Optimizing multiway joins in a map-reduce environment.
\newblock {\em {IEEE} Trans. Knowl. Data Eng.}, 23(9):1282--1298, 2011.

\bibitem{agarwal14}
A.~Agarwal, O.~Chapelle, M.~Dudik, and J.~Langford.
\newblock A reliable effective terascale linear learning system.
\newblock {\em Journal of Machine Learning Research}, 15:1111--1133, 2014.

\bibitem{A09}
R.~Agrawal, S.~Gollapudi, A.~Halverson, and S.~Ieong.
\newblock Diversifying search results.
\newblock In {\em International Conference on Web Search and Web Data Mining
  (WSDM)}, pages 5--14. {ACM}, 2009.

\bibitem{akdere11}
M.~Akdere, U.~Cetintemel, M.~Riondato, E.~Upfal, and S.~B. Zdonik.
\newblock The case for predictive database systems: Opportunities and
  challenges.
\newblock In {\em Conference on Innovative Data Systems Research (CIDR)}, pages
  167--174. www.cidrdb.org, 2011.

\bibitem{DBLP:conf/icalp/AmarilliBS15}
A.~Amarilli, P.~Bourhis, and P.~Senellart.
\newblock Provenance circuits for trees and treelike instances.
\newblock In {\em International Colloquium on Automata, Languages, and
  Programming (ICALP)}, volume 9135 of {\em LNCS}, pages 56--68. Springer,
  2015.

\bibitem{AmelootGKNS15}
T.~J. Ameloot, G.~Geck, B.~Ketsman, F.~Neven, and T.~Schwentick.
\newblock Parallel-correctness and transferability for conjunctive queries.
\newblock In {\em Proceedings of the 34th {ACM} Symposium on Principles of
  Database Systems, {PODS} 2015}, pages 47--58, 2015.

\bibitem{compas}
J.~Angwin, J.~Larson, S.~Mattu, and L.~Kirchner.
\newblock Machine bias.
\newblock ProPublica, May 2016.
\newblock
  \url{https://www.propublica.org/article/machine-bias-risk-assessments-in-criminal-sentencing}.

\bibitem{DBLP:conf/sigmod/ArefCGKOPVW15}
M.~Aref, B.~ten Cate, T.~J. Green, B.~Kimelfeld, D.~Olteanu, E.~Pasalic, T.~L.
  Veldhuizen, and G.~Washburn.
\newblock Design and implementation of the {L}ogic{B}lox system.
\newblock In {\em International Conference on Management of Data (SIGMOD)},
  pages 1371--1382. ACM, 2015.

\bibitem{ABLM14}
M.~Arenas, P.~Barcel\'o, L.~Libkin, and F.~Murlak.
\newblock {\em {Foundations of Data Exchange}}.
\newblock Cambridge University Press, 2014.

\bibitem{DBLP:conf/pods/ArenasGP14}
M.~Arenas, G.~Gottlob, and A.~Pieris.
\newblock Expressive languages for querying the semantic web.
\newblock In {\em Symposium on Principles of Database Systems (PODS)}, pages
  14--26. ACM, 2014.

\bibitem{DBLP:journals/ws/ArenasGKMZ16}
M.~Arenas, B.~C. Grau, E.~Kharlamov, S.~Marciuska, and D.~Zheleznyakov.
\newblock Faceted search over {RDF}-based knowledge graphs.
\newblock {\em J. Web Sem.}, 37:55--74, 2016.

\bibitem{AMRV16}
M.~Arenas, F.~Maturana, C.~Riveros, and D.~Vrgoc.
\newblock A framework for annotating {CSV}-like data.
\newblock {\em Proceedings of the VLDB Endowment}, 9(11), 2016.

\bibitem{AKRZ14}
A.~Artale, R.~Kontchakov, V.~Ryzhikov, and M.~Zakharyaschev.
\newblock A cookbook for temporal conceptual data modelling with description
  logics.
\newblock {\em ACM Trans.\ on Computational Logic}, 15(3):25:1--25:50, 2014.

\bibitem{atserias2008size}
A.~Atserias, M.~Grohe, and D.~Marx.
\newblock Size bounds and query plans for relational joins.
\newblock {\em {SIAM} J. Comput.}, 42(4):1737--1767, 2013.

\bibitem{BaKP12}
F.~Baader, M.~Knechtel, and R.~Pe{\~{n}}aloza.
\newblock Context-dependent views to axioms and consequences of semantic web
  ontologies.
\newblock {\em J. Web Sem.}, 12:22--40, 2012.

\bibitem{BLMS11}
J.-F. Baget, M.~Lecl{\`e}re, M.-L. Mugnier, and E.~Salvat.
\newblock On rules with existential variables: Walking the decidability line.
\newblock {\em Artificial Intelligence}, 175(9--10):1620--1654, 2011.

\bibitem{Milo-et-al:querying-process-instances}
E.~Balan, T.~Milo, and T.~Sterenzy.
\newblock {BP-Ex}: a uniform query engine for business process execution
  traces.
\newblock In {\em International Conference on Extending Database Technology
  (EDBT)}, pages 713--716. {ACM}, 2010.

\bibitem{DBLP:conf/icdt/BaranyCKOV16}
V.~B{\'{a}}r{\'{a}}ny, B.~ten Cate, B.~Kimelfeld, D.~Olteanu, and Z.~Vagena.
\newblock Declarative probabilistic programming with datalog.
\newblock In {\em International Conference on Database Theory (ICDT)},
  volume~48 of {\em LIPIcs}, pages 7:1--7:19. Schloss Dagstuhl--LZI, 2016.

\bibitem{BarocasSelbst}
S.~Barocas and A.~D. Selbst.
\newblock Big data's disparate impact.
\newblock {\em California Law Review}, 104, 2016.

\bibitem{beame13:_commun}
P.~Beame, P.~Koutris, and D.~Suciu.
\newblock Communication steps for parallel query processing.
\newblock In {\em Symposium on Principles of Database Systems (PODS)}, pages
  273--284. ACM, 2013.

\bibitem{DBLP:journals/jacm/BenediktFG08}
M.~Benedikt, W.~Fan, and F.~Geerts.
\newblock {XP}ath satisfiability in the presence of {DTDs}.
\newblock {\em J. {ACM}}, 55(2), 2008.

\bibitem{bengio13}
Y.~Bengio, A.~C. Courville, and P.~Vincent.
\newblock Representation learning: {A} review and new perspectives.
\newblock {\em {IEEE} Transactions on Pattern Analysis and Machine
  Intelligence}, 35(8):1798--1828, 2013.

\bibitem{bertossi-book}
L.~Bertossi.
\newblock {\em Database Repairing and Consistent Query Answering}.
\newblock Morgan\&Claypool Publishers, 2011.

\bibitem{DBLP:journals/tods/BexNSV10}
G.~J. Bex, F.~Neven, T.~Schwentick, and S.~Vansummeren.
\newblock Inference of concise regular expressions and {DTDs}.
\newblock {\em {ACM} Trans. Database Syst.}, 35(2), 2010.

\bibitem{Bhattacharya-et-al-BPM-2007}
K.~Bhattacharya, C.~Gerede, R.~Hull, R.~Liu, and J.~Su.
\newblock Towards formal analysis of artifact-centric business process models.
\newblock In {\em International Conference on Business Process Management
  (BPM)}, volume 4714 of {\em LNCS}, pages 288--304. Springer, 2007.

\bibitem{BiOr15}
M.~Bienvenu and M.~Ortiz.
\newblock Ontology-mediated query answering with data-tractable description
  logics.
\newblock In {\em International Summer School on Reasoning Web}, volume 9203 of
  {\em LNCS}, pages 218--307. Springer, 2015.

\bibitem{BCLW14}
M.~Bienvenu, B.~ten Cate, C.~Lutz, and F.~Wolter.
\newblock Ontology-based data access: {A} study through {Disjunctive}
  {Datalog}, {CSP}, and {MMSNP}.
\newblock {\em {ACM} Trans. Database Syst.}, 39(4):33:1--33:44, 2014.

\bibitem{BoDP15}
S.~Borgwardt, F.~Distel, and R.~Pe{\~n}aloza.
\newblock The limits of decidability in fuzzy description logics with general
  concept inclusions.
\newblock {\em Artificial Intelligence}, 218:23--55, 2015.

\bibitem{DBLP:conf/webdb/CafarellaSE07}
M.~J. Cafarella, D.~Suciu, and O.~Etzioni.
\newblock Navigating extracted data with schema discovery.
\newblock In {\em International Workshop on the Web and Databases (WebDB)},
  2007.

\bibitem{DBLP:conf/sigmod/CaiVPAHJ13}
Z.~Cai, Z.~Vagena, L.~L. Perez, S.~Arumugam, P.~J. Haas, and C.~M. Jermaine.
\newblock Simulation of database-valued markov chains using simsql.
\newblock In {\em International Conference on Management of Data (SIGMOD)},
  pages 637--648. {ACM}, 2013.

\bibitem{CGLLR07}
D.~Calvanese, G.~{De Giacomo}, D.~Lembo, M.~Lenzerini, and R.~Rosati.
\newblock Tractable reasoning and efficient query answering in description
  logics: The \emph{DL-Lite} family.
\newblock {\em J. Autom. Reasoning}, 39(3):385--429, 2007.

\bibitem{CaDL08}
D.~Calvanese, G.~De~Giacomo, and M.~Lenzerini.
\newblock Conjunctive query containment and answering under description logics
  constraints.
\newblock {\em ACM Trans.\ on Computational Logic}, 9(3):22.1--22.31, 2008.

\bibitem{Calvanese-et-al-PODS-2013}
D.~Calvanese, G.~{De Giacomo}, and M.~Montali.
\newblock Foundations of data-aware process analysis: a database theory
  perspective.
\newblock In {\em Symposium on Principles of Database Systems (PODS)}, pages
  1--12. ACM, 2013.

\bibitem{DBLP:journals/pvldb/CebiricGM15}
S.~Cebiric, F.~Goasdou{\'{e}}, and I.~Manolescu.
\newblock Query-oriented summarization of {RDF} graphs.
\newblock {\em {Proceedings of the VLDB Endowment}}, 8(12):2012--2015, 2015.

\bibitem{choromanska15}
A.~Choromanska and J.~Langford.
\newblock Logarithmic time online multiclass prediction.
\newblock In {\em Advances in Neural Information Processing Systems (NIPS)},
  2015.

\bibitem{chu15:_from}
S.~Chu, M.~Balazinska, and D.~Suciu.
\newblock From theory to practice: Efficient join query evaluation in a
  parallel database system.
\newblock In {\em International Conference on Management of Data (SIGMOD)},
  pages 63--78. ACM, 2015.

\bibitem{cisse13}
M.~Ciss\'e, N.~Usunier, T.~Artieres, and P.~Gallinari.
\newblock Robust {B}loom filters for large multilabel classification tasks.
\newblock In {\em Advances in Neural Information Processing Systems (NIPS)},
  2013.

\bibitem{codd75}
E.~F. Codd.
\newblock Understanding relations (installment \#7).
\newblock {\em FDT - Bulletin of ACM SIGMOD}, 7(3):23--28, 1975.

\bibitem{cohen_et_al:LIPIcs:2015:4981}
S.~Cohen and Y.~Y. Weiss.
\newblock {Learning Tree Patterns from Example Graphs}.
\newblock In {\em International Conference on Database Theory (ICDT)},
  volume~31 of {\em LIPIcs}, pages 127--143. Schloss Dagstuhl--LZI, 2015.

\bibitem{CK14}
V.~Cortier and S.~Kremer.
\newblock Formal models and techniques for analyzing security protocols: {A}
  tutorial.
\newblock {\em Foundations and Trends in Programming Languages}, 1(3):151--267,
  2014.

\bibitem{DBLP:conf/pods/CzerwinskiMPP15}
W.~Czerwinski, W.~Martens, P.~Parys, and M.~Przybylko.
\newblock The (almost) complete guide to tree pattern containment.
\newblock In {\em Symposium on Principles of Database Systems (PODS)}, pages
  117--130. ACM, 2015.

\bibitem{DBLP:journals/jacm/DalviS12}
N.~N. Dalvi and D.~Suciu.
\newblock The dichotomy of probabilistic inference for unions of conjunctive
  queries.
\newblock {\em J. {ACM}}, 59(6):30, 2012.

\bibitem{Danaher}
J.~Danaher.
\newblock The threat of algocracy: Reality, resistance and accommodation.
\newblock {\em Philosophy and Technology}, forthcoming.

\bibitem{date2005}
C.~J. Date.
\newblock {\em Database in Depth -- Relational Theory for Practitioners}.
\newblock O'Reilly, 2005.

\bibitem{DattaSZ16}
A.~Datta, S.~Sen, and Y.~Zick.
\newblock Algorithmic transparency via quantitative input influence: Theory and
  experiments with learning systems.
\newblock In {\em Symposium on Security and Privacy ({SP})}, pages 598--617.
  IEEE, 2016.

\bibitem{DBLP:journals/popets/DattaTD15}
A.~Datta, M.~C. Tschantz, and A.~Datta.
\newblock Automated experiments on ad privacy settings.
\newblock {\em PoPETs}, 2015(1):92--112, 2015.

\bibitem{davidson-freire:sci-wf:sigmod-2008}
S.~B. Davidson and J.~Freire.
\newblock Provenance and scientific workflows: {Challenges} and opportunities.
\newblock In {\em International Conference on Management of Data (SIGMOD)},
  pages 1345--1350. ACM, 2008.

\bibitem{dayal-ETL-EDBT-invited-2009}
U.~Dayal, M.~Castellanos, A.~Simitsis, and K.~Wilkinson.
\newblock Data integration flows for business intelligence.
\newblock In {\em International Conference on Extending Database Technology
  (EDBT)}, pages 1--11. {ACM}, 2009.

\bibitem{dembczynski10}
K.~Dembczynski, W.~Cheng, and E.~H\"ullermeier.
\newblock Bayes optimal multilabel classification via probabilistic classifier
  chains.
\newblock In {\em International Conference on Machine Learning (ICML)}, pages
  279--286. Omnipress, 2010.

\bibitem{Milo:PODS-2011}
D.~Deutch and T.~Milo.
\newblock A quest for beauty and wealth (or, business processes for database
  researchers).
\newblock In {\em Symposium on Principles of Database Systems (PODS)}, pages
  1--12. ACM, 2011.

\bibitem{DHPV-analysis-et-al:ICDT-09}
A.~Deutsch, R.~Hull, F.~Patrizi, and V.~Vianu.
\newblock Automatic verification of data-centric business processes.
\newblock In {\em International Conference on Database Theory (ICDT)}. {ACM},
  2009.

\bibitem{DeutschHV14}
A.~Deutsch, R.~Hull, and V.~Vianu.
\newblock Automatic verification of database-centric systems.
\newblock {\em {SIGMOD} Record}, 43(3):5--17, 2014.

\bibitem{domingos00}
P.~M. Domingos and G.~Hulten.
\newblock Mining high-speed data streams.
\newblock In {\em International Conference on Knowledge Discovery and Data
  Mining (KDD)}, pages 71--80. {ACM}, 2000.

\bibitem{DrLB15}
Z.~Dragisic, P.~Lambrix, and E.~Blomqvist.
\newblock Integrating ontology debugging and matching into the {eXtreme} design
  methodology.
\newblock In {\em Workshop on Ontology and Semantic Web Patterns (WOP)}, volume
  1461 of {\em CEUR Workshop Proceedings}, 2015.

\bibitem{DBLP:journals/pvldb/DrosouP12}
M.~Drosou and E.~Pitoura.
\newblock {DisC} diversity: result diversification based on dissimilarity and
  coverage.
\newblock {\em {Proceedings of the VLDB Endowment}}, 6(1):13--24, 2012.

\bibitem{D11}
C.~Dwork.
\newblock A firm foundation for private data analysis.
\newblock {\em Commun. {ACM}}, 54(1):86--95, 2011.

\bibitem{Detal12}
C.~Dwork, M.~Hardt, T.~Pitassi, O.~Reingold, and R.~S. Zemel.
\newblock Fairness through awareness.
\newblock In {\em Innovations in Theoretical Computer Science (ITCS)}, pages
  214--226. {ACM}, 2012.

\bibitem{eiter-kr16}
T.~Eiter, T.~Lukasiewicz, and L.~Predoiu.
\newblock Generalized consistent query answering under existential rules.
\newblock In {\em International Conference on Principles of Knowledge
  Representation and Reasoning (KR)}, pages 359--368. {AAAI} Press, 2016.

\bibitem{ElKS06}
C.~Elsenbroich, O.~Kutz, and U.~Sattler.
\newblock A case for abductive reasoning over ontologies.
\newblock In {\em International Workshop on OWL (OWLED)}, volume 216 of {\em
  CEUR Workshop Proceedings}, 2006.

\bibitem{DBLP:conf/www/EnglehardtREZMN15}
S.~Englehardt, D.~Reisman, C.~Eubank, P.~Zimmerman, J.~Mayer, A.~Narayanan, and
  E.~W. Felten.
\newblock Cookies that give you away: The surveillance implications of web
  tracking.
\newblock In {\em International Conference on World Wide Web (WWW)}, pages
  289--299. {ACM}, 2015.

\bibitem{FaginKRV14}
R.~Fagin, B.~Kimelfeld, F.~Reiss, and S.~Vansummeren.
\newblock Cleaning inconsistencies in information extraction via prioritized
  repairs.
\newblock In {\em Symposium on Principles of Database Systems (PODS)}. ACM,
  2014.

\bibitem{fagin2015document}
R.~Fagin, B.~Kimelfeld, F.~Reiss, and S.~Vansummeren.
\newblock Document spanners: A formal approach to information extraction.
\newblock {\em J. ACM}, 62(2):12, 2015.

\bibitem{FMSSS08}
J.~Feldman, S.~Muthukrishnan, A.~Sidiropoulos, C.~Stein, and Z.~Svitkina.
\newblock On distributing symmetric streaming computations.
\newblock In {\em Symposium on Discrete Algorithms (SODA)}, pages 710--719.
  {SIAM}, 2008.

\bibitem{FGTT11}
E.~Franconi, P.~Guagliardo, M.~Trevisan, and S.~Tessaris.
\newblock {Quelo}: an ontology-driven query interface.
\newblock In {\em Workshop on Description Logics (DL)}, volume 745 of {\em CEUR
  Workshop Proceedings}, 2011.

\bibitem{FBetal10}
B.~C.~M. Fung, K.~Wang, R.~Chen, and P.~S. Yu.
\newblock Privacy-preserving data publishing: {A} survey of recent
  developments.
\newblock {\em {ACM} Comput. Surv.}, 42(4), 2010.

\bibitem{DBLP:conf/popl/Goodman13}
N.~D. Goodman.
\newblock The principles and practice of probabilistic programming.
\newblock In {\em Symposium on Principles of Programming Languages (POPL)},
  pages 399--402. ACM, 2013.

\bibitem{GKKP*14}
G.~Gottlob, S.~Kikot, R.~Kontchakov, V.~V. Podolskii, T.~Schwentick, and
  M.~Zakharyaschev.
\newblock The price of query rewriting in ontology-based data access.
\newblock {\em Artificial Intelligence}, 213:42--59, 2014.

\bibitem{DBLP:journals/tods/GottlobKP05}
G.~Gottlob, C.~Koch, and R.~Pichler.
\newblock Efficient algorithms for processing {XP}ath queries.
\newblock {\em {ACM} Trans. Database Syst.}, 30(2):444--491, 2005.

\bibitem{GoOP14}
G.~Gottlob, G.~Orsi, and A.~Pieris.
\newblock Query rewriting and optimization for ontological databases.
\newblock {\em {ACM} Trans. Database Syst.}, 39(3):25:1--25:46, 2014.

\bibitem{DBLP:journals/jacm/GottlobS10}
G.~Gottlob and P.~Senellart.
\newblock Schema mapping discovery from data instances.
\newblock {\em J. {ACM}}, 57(2), 2010.

\bibitem{GrozMR15}
B.~Groz, T.~Milo, and S.~Roy.
\newblock On the complexity of evaluating order queries with the crowd.
\newblock {\em {IEEE} Data Eng. Bull.}, 38(3):44--58, 2015.

\bibitem{haas99:_rippl}
P.~J. Haas and J.~M. Hellerstein.
\newblock Ripple joins for online aggregation.
\newblock In {\em International Conference on Management of Data (SIGMOD)},
  pages 287--298. ACM, 1999.

\bibitem{DBLP:conf/kdd/HajianBC16}
S.~Hajian, F.~Bonchi, and C.~Castillo.
\newblock Algorithmic bias: From discrimination discovery to fairness-aware
  data mining.
\newblock In {\em International Conference on Knowledge Discovery and Data
  Mining (KDD)}, pages 2125--2126. ACM, 2016.

\bibitem{hellerstein97:_onlin}
J.~M. Hellerstein, P.~J. Haas, and H.~J. Wang.
\newblock Online aggregation.
\newblock In {\em International Conference on Management of Data (SIGMOD)},
  pages 171--182. ACM, 1997.

\bibitem{Hepp15}
M.~Hepp.
\newblock The web of data for e-commerce: {Schema.org} and {GoodRelations} for
  researchers and practitioners.
\newblock In {\em International Conference on Web Engineering (ICWE)}, volume
  9114 of {\em LNCS}, pages 723--727. Springer, 2015.

\bibitem{hu16:_towar}
X.~Hu and K.~Yi.
\newblock Towards a worst-case i/o-optimal algorithm for acyclic joins.
\newblock In {\em Symposium on Principles of Database Systems (PODS)}. ACM,
  2016.

\bibitem{Hull-Su-NSF-report-2009}
R.~Hull and J.~Su.
\newblock {NSF Workshop on Data-Centric Workflows}, May, 2009.
\newblock \url{http://dcw2009.cs.ucsb.edu/report.pdf}.

\bibitem{IL84}
T.~Imielinski and W.~Lipski.
\newblock Incomplete information in relational databases.
\newblock {\em J. ACM}, 31(4):761--791, 1984.

\bibitem{jasinska16}
K.~Jasinska, K.~Dembczynski, , R.~Busa-Fekete, K.~Pfannschmidt, T.~Klerx, and
  E.~H\"ullermeier.
\newblock Extreme {F}-measure maximization using sparse probability estimates.
\newblock In {\em International Conference on Machine Learning (ICML)}.
  JMLR.org, 2016.

\bibitem{journals/pvldb/JhaS12}
A.~K. Jha and D.~Suciu.
\newblock Probabilistic databases with {MarkoViews}.
\newblock {\em Proceedings of the VLDB Endowment}, 5(11):1160--1171, 2012.

\bibitem{DBLP:conf/icdt/KaminskiK16}
M.~Kaminski and E.~V. Kostylev.
\newblock Beyond well-designed {SPARQL}.
\newblock In {\em International Conference on Database Theory (ICDT)},
  volume~48 of {\em LIPIcs}, pages 5:1--5:18. Schloss Dagstuhl -- LZI, 2016.

\bibitem{DBLP:journals/tvcg/KandelPHH12}
S.~Kandel, A.~Paepcke, J.~M. Hellerstein, and J.~Heer.
\newblock Enterprise data analysis and visualization: An interview study.
\newblock {\em {IEEE} Trans. Vis. Comput. Graph.}, 18(12):2917--2926, 2012.

\bibitem{DBLP:conf/pods/Kimelfeld12}
B.~Kimelfeld.
\newblock A dichotomy in the complexity of deletion propagation with functional
  dependencies.
\newblock In {\em Symposium on Principles of Database Systems (PODS)}, pages
  191--202. {ACM}, 2012.

\bibitem{DBLP:journals/tods/KimelfeldK14}
B.~Kimelfeld and P.~G. Kolaitis.
\newblock The complexity of mining maximal frequent subgraphs.
\newblock {\em {ACM} Trans. Database Syst.}, 39(4):32:1--32:33, 2014.

\bibitem{DBLP:journals/jacm/KimelfeldR14}
B.~Kimelfeld and C.~R{\'{e}}.
\newblock Transducing {Markov} sequences.
\newblock {\em J. {ACM}}, 61(5):32:1--32:48, 2014.

\bibitem{DBLP:series/sfsc/KimelfeldS13}
B.~Kimelfeld and P.~Senellart.
\newblock Probabilistic {XML:} models and complexity.
\newblock In {\em Advances in Probabilistic Databases for Uncertain Information
  Management}, volume 304 of {\em Studies in Fuzziness and Soft Computing},
  pages 39--66. Springer, 2013.

\bibitem{KlGu10}
S.~Klarman and V.~Guti{\'e}rrez-Basulto.
\newblock $\mathcal{ALC}_{\mathcal{alc}}$: A context description logic.
\newblock In {\em European Conference on Logics in Artificial Intelligence
  (JELIA)}, volume 6341 of {\em LNCS}, pages 208--220. Springer, 2010.

\bibitem{KLTWZ11}
R.~Kontchakov, C.~Lutz, D.~Toman, F.~Wolter, and M.~Zakharyaschev.
\newblock The combined approach to ontology-based data access.
\newblock In {\em International Joint Conference on Artificial Intelligence
  (IJCAI)}, pages 2656--2661. {IJCAI/AAAI}, 2011.

\bibitem{KBS16}
P.~Koutris, P.~Beame, and D.~Suciu.
\newblock Worst-case optimal algorithms for parallel query processing.
\newblock In {\em International Conference on Database Theory (ICDT)},
  volume~48 of {\em LIPIcs}, pages 8:1--8:18. Schloss Dagstuhl -- LZI, 2016.

\bibitem{DBLP:conf/pods/KoutrisW15}
P.~Koutris and J.~Wijsen.
\newblock The data complexity of consistent query answering for self-join-free
  conjunctive queries under primary key constraints.
\newblock In {\em Symposium on Principles of Database Systems (PODS)}, pages
  17--29. {ACM}, 2015.

\bibitem{Kroll2017}
J.~A. Kroll, J.~Huey, S.~Barocas, E.~W. Felten, J.~R. Reidenberg, D.~G.
  Robinson, and H.~Yu.
\newblock Accountable algorithms.
\newblock {\em University of Pennsylvania Law Review}, 165, 2017.

\bibitem{Letal14}
M.~L{\'{e}}cuyer, G.~Ducoffe, F.~Lan, A.~Papancea, T.~Petsios, R.~Spahn,
  A.~Chaintreau, and R.~Geambasu.
\newblock {XR}ay: Enhancing the web's transparency with differential
  correlation.
\newblock In {\em {USENIX} Security Symposium}, pages 49--64. {USENIX}
  Association, 2014.

\bibitem{LMRST15}
D.~Lembo, J.~Mora, R.~Rosati, D.~F. Savo, and E.~Thorstensen.
\newblock Mapping analysis in ontology-based data access: Algorithms and
  complexity.
\newblock In {\em International Semantic Web Conference (ISWC)}, volume 9366 of
  {\em LNCS}, pages 217--234. Springer, 2015.

\bibitem{Lenzerini02}
M.~Lenzerini.
\newblock {Data integration: a theoretical perspective}.
\newblock In {\em ACM Symposium on Principles of Database Systems (PODS)},
  pages 233--246. ACM, 2002.

\bibitem{lerman2013big}
J.~Lerman.
\newblock Big data and its exclusions.
\newblock {\em Stanford Law Review Online}, 66, 2013.

\bibitem{li16:_wander}
F.~Li, B.~Wu, K.~Yi, and Z.~Zhao.
\newblock Wander join: Online aggregation via random walks.
\newblock In {\em International Conference on Management of Data (SIGMOD)},
  pages 615--629. ACM, 2016.

\bibitem{pods14}
L.~Libkin.
\newblock Incomplete information: what went wrong and how to fix it.
\newblock In {\em Symposium on Principles of Database Systems (PODS)}, pages
  1--13. ACM, 2014.

\bibitem{DBLP:journals/ai/Libkin16}
L.~Libkin.
\newblock Certain answers as objects and knowledge.
\newblock {\em Artificial Intelligence}, 232:1--19, 2016.

\bibitem{Ltods16}
L.~Libkin.
\newblock {SQL}'s three-valued logic and certain answers.
\newblock {\em {ACM} Trans. Database Syst.}, 41(1):1, 2016.

\bibitem{Lipski-certain-answers}
W.~Lipski.
\newblock On semantic issues connected with incomplete information databases.
\newblock {\em {ACM} Trans. Database Syst.}, 4(3):262--296, 1979.

\bibitem{Liu-Vaculin:artifact-centric-monitoring}
R.~Liu, R.~Vacul{\'{\i}}n, Z.~Shan, A.~Nigam, and F.~Y. Wu.
\newblock Business artifact-centric modeling for real-time performance
  monitoring.
\newblock In {\em International Conference on Business Process Management
  (BPM)}, pages 265--280, 2011.

\bibitem{GSM-and-CMMN:2012}
M.~Marin, R.~Hull, and R.~Vacul\'{\i}n.
\newblock {Data-centric BPM and the emerging Case Management standard: A short
  survey}.
\newblock In {\em Business Process Management Workshops}, pages 24--30, 2012.

\bibitem{DBLP:conf/www/MartensNV15}
W.~Martens, F.~Neven, and S.~Vansummeren.
\newblock {SCULPT:} {A} schema language for tabular data on the web.
\newblock In {\em International Conference on World Wide Web (WWW)}, pages
  702--720. {ACM}, 2015.

\bibitem{Metal15}
V.~Z. Moffitt, J.~Stoyanovich, S.~Abiteboul, and G.~Miklau.
\newblock Collaborative access control in {W}ebdam{L}og.
\newblock In {\em International Conference on Management of Data (SIGMOD)},
  pages 197--211. ACM, 2015.

\bibitem{morales15}
G.~D.~F. Morales and A.~Bifet.
\newblock {SAMOA}: {S}calable advanced massive online analysis.
\newblock {\em Journal of Machine Learning Research}, 16:149--153, 2015.

\bibitem{WHMay2016}
C.~Mu{\~n}oz, M.~Smith, and D.~Patil.
\newblock Big data: A report on algorithmic systems, opportunity, and civil
  rights.
\newblock {\em Executive Office of the President, The White House}, May 2016.

\bibitem{ngo2012worst}
H.~Q. Ngo, E.~Porat, C.~R{\'{e}}, and A.~Rudra.
\newblock Worst-case optimal join algorithms: [extended abstract].
\newblock In {\em Symposium on Principles of Database Systems (PODS)}, pages
  37--48. ACM, 2012.

\bibitem{NgOS16}
N.~Ngo, M.~Ortiz, and M.~Simkus.
\newblock Closed predicates in description logics: Results on combined
  complexity.
\newblock In {\em International Conference on the Principles of Knowledge
  Representation and Reasoning (KR)}, pages 237--246. AAAI Press, 2016.

\bibitem{Nigam-Caswell-2003}
A.~Nigam and N.~Caswell.
\newblock {Business Artifacts: An Approach to Operational Specification}.
\newblock {\em IBM Systems Journal}, 42(3), 2003.

\bibitem{DBLP:conf/vlds/NiuZRS12}
F.~Niu, C.~Zhang, C.~Re, and J.~W. Shavlik.
\newblock {DeepDive}: Web-scale knowledge-base construction using statistical
  learning and inference.
\newblock In {\em International Workshop on Searching and Integrating New Web
  Data Sources}, volume 884 of {\em CEUR Workshop Proceedings}, pages 25--28.
  CEUR-WS.org, 2012.

\bibitem{DBLP:journals/tods/OlteanuZ15}
D.~Olteanu and J.~Z{\'{a}}vodn{\'{y}}.
\newblock Size bounds for factorised representations of query results.
\newblock {\em {ACM} Trans. Database Syst.}, 40(1):2, 2015.

\bibitem{DBLP:journals/jcss/PapadimitriouY99}
C.~H. Papadimitriou and M.~Yannakakis.
\newblock On the complexity of database queries.
\newblock {\em J. Comput. Syst. Sci.}, 58(3):407--427, 1999.

\bibitem{DBLP:conf/www/PezoaRSUV16}
F.~Pezoa, J.~L. Reutter, F.~Suarez, M.~Ugarte, and D.~Vrgoc.
\newblock Foundations of {JSON} schema.
\newblock In {\em International Conference on World Wide Web (WWW)}, pages
  263--273. {ACM}, 2016.

\bibitem{PLCD*08}
A.~Poggi, D.~Lembo, D.~Calvanese, G.~De~Giacomo, M.~Lenzerini, and R.~Rosati.
\newblock Linking data to ontologies.
\newblock {\em J.\ on Data Semantics}, X:133--173, 2008.

\bibitem{prabhu14}
Y.~Prabhu and M.~Varma.
\newblock {FastXML}: a fast, accurate and stable tree-classifier for extreme
  multi-label learning.
\newblock In {\em International Conference on Knowledge Discovery and Data
  Mining (KDD)}, pages 263--272. ACM, 2014.

\bibitem{riondato11}
M.~Riondato, M.~Akdere, U.~Cetintemel, S.~B. Zdonik, and E.~Upfal.
\newblock The vc-dimension of {SQL} queries and selectivity estimation through
  sampling.
\newblock In {\em European Conference on Machine Learning and Knowledge
  Discovery in Databases (ECML/PKDD)}, volume 6912 of {\em LNCS}, pages
  661--676. Springer, 2011.

\bibitem{salakhutdinov09}
R.~Salakhutdinov and G.~E. Hinton.
\newblock Semantic hashing.
\newblock {\em Int. Journal of Approximate Reasoning}, 50(7):969--978, 2009.

\bibitem{SarmaPW16}
A.~D. Sarma, A.~G. Parameswaran, and J.~Widom.
\newblock Towards globally optimal crowdsourcing quality management: The
  uniform worker setting.
\newblock In {\em International Conference on Management of Data (SIGMOD)},
  pages 47--62, 2016.

\bibitem{DBLP:conf/sigmod/SchleichOC16}
M.~Schleich, D.~Olteanu, and R.~Ciucanu.
\newblock Learning linear regression models over factorized joins.
\newblock In {\em International Conference on Management of Data (SIGMOD)},
  pages 3--18. ACM, 2016.

\bibitem{scho_lw}
B.~Sch\"olkopf and A.~Smola.
\newblock {\em Learning with Kernels: Support Vector Machines, Regularization,
  Optimization, and Beyond}.
\newblock MIT Press, 2001.

\bibitem{DBLP:conf/sigmod/SelingerACLP79}
P.~G. Selinger, M.~M. Astrahan, D.~D. Chamberlin, R.~A. Lorie, and T.~G. Price.
\newblock Access path selection in a relational database management system.
\newblock In {\em International Conference on Management of Data (SIGMOD)},
  pages 23--34. {ACM}, 1979.

\bibitem{DBLP:conf/vldb/ShenDNR07}
W.~Shen, A.~Doan, J.~F. Naughton, and R.~Ramakrishnan.
\newblock Declarative information extraction using datalog with embedded
  extraction predicates.
\newblock In {\em International Conference on Very Large Data Bases (VLDB)},
  pages 1033--1044. ACM, 2007.

\bibitem{DBLP:journals/pvldb/ShinWWSZR15}
J.~Shin, S.~Wu, F.~Wang, C.~D. Sa, C.~Zhang, and C.~R{\'{e}}.
\newblock Incremental knowledge base construction using deepdive.
\newblock {\em {Proceedings of the VLDB Endowment}}, 8(11):1310--1321, 2015.

\bibitem{DBLP:conf/icdt/StaworkoBGHPS15}
S.~Staworko, I.~Boneva, J.~E.~L. Gayo, S.~Hym, E.~G. Prud'hommeaux, and H.~R.
  Solbrig.
\newblock Complexity and expressiveness of shex for {RDF}.
\newblock In {\em International Conference on Database Theory (ICDT)},
  volume~31 of {\em LIPIcs}, pages 195--211. Schloss Dagstuhl -- LZI, 2015.

\bibitem{StaworkoCM12}
S.~Staworko, J.~Chomicki, and J.~Marcinkowski.
\newblock Prioritized repairing and consistent query answering in relational
  databases.
\newblock {\em Ann. Math. Artif. Intell.}, 64(2-3):209--246, 2012.

\bibitem{DBLP:conf/edbt/StoyanovichAM16}
J.~Stoyanovich, S.~Abiteboul, and G.~Miklau.
\newblock Data responsibly: Fairness, neutrality and transparency in data
  analysis.
\newblock In {\em International Conference on Extending Database Technology
  (EDBT)}, pages 718--719. OpenProceedings.org, 2016.

\bibitem{probabilistic-book}
D.~Suciu, D.~Olteanu, C.~Re, and C.~Koch.
\newblock {\em {Probabilistic Databases}}.
\newblock Morgan\&Claypool Publishers, 2011.

\bibitem{ST94}
D.~Suciu and V.~Tannen.
\newblock A query language for {NC}.
\newblock In {\em Symposium on Principles of Database Systems (PODS)}, pages
  167--178. ACM, 1994.

\bibitem{Sun-Su-Yang-TMIS-2016}
Y.~Sun, J.~Su, and J.~Yang.
\newblock Universal artifacts.
\newblock {\em ACM Trans. on Management Information Systems}, 7(1), 2016.

\bibitem{DBLP:journals/cacm/Sweeney13}
L.~Sweeney.
\newblock Discrimination in online ad delivery.
\newblock {\em Commun. {ACM}}, 56(5):44--54, 2013.

\bibitem{DBLP:journals/tods/CateDK13}
B.~ten Cate, V.~Dalmau, and P.~G. Kolaitis.
\newblock Learning schema mappings.
\newblock {\em {ACM} Trans. Database Syst.}, 38(4):28, 2013.

\bibitem{DBLP:journals/corr/TramerAGHHHJL15}
F.~Tram{\`{e}}r, V.~Atlidakis, R.~Geambasu, D.~J. Hsu, J.~Hubaux, M.~Humbert,
  A.~Juels, and H.~Lin.
\newblock Discovering unwarranted associations in data-driven applications with
  the fairtest testing toolkit.
\newblock {\em CoRR}, abs/1510.02377, 2015.

\bibitem{Tschorsch-Scheuermann-Blockchain}
F.~Tschorsch and B.~Scheuermann.
\newblock Bitcoin and beyond: A technical survey on decentralized digital
  currencies.
\newblock Cryptology ePrint Archive, Report 2015/464, 2015.
\newblock \url{http://eprint.iacr.org/}.

\bibitem{valiant84}
L.~Valiant.
\newblock A theory of the learnable.
\newblock {\em Commun. ACM}, 17(11):1134--1142, 1984.

\bibitem{valiant90}
L.~G. Valiant.
\newblock A bridging model for parallel computation.
\newblock {\em Commun. {ACM}}, 33(8):103--111, 1990.

\bibitem{DBLP:conf/stoc/Vardi82}
M.~Y. Vardi.
\newblock The complexity of relational query languages (extended abstract).
\newblock In {\em Symposium on Theory of Computing (STOC)}, pages 137--146.
  {ACM}, 1982.

\bibitem{veldhuizen14}
T.~L. Veldhuizen.
\newblock Triejoin: {A} simple, worst-case optimal join algorithm.
\newblock In {\em International Conference on Database Theory (ICDT)}, pages
  96--106. OpenProceedings.org, 2014.

\bibitem{weinberger09}
K.~Weinberger, A.~Dasgupta, J.~Langford, A.~Smola, and J.~Attenberg.
\newblock Feature hashing for large scale multitask learning.
\newblock In {\em International Conference on Machine Learning (ICML)}, pages
  1113--1120. {ACM}, 2009.

\bibitem{DBLP:journals/pvldb/Wu0HZL15}
T.~Wu, L.~Chen, P.~Hui, C.~J. Zhang, and W.~Li.
\newblock Hear the whole story: Towards the diversity of opinion in
  crowdsourcing markets.
\newblock {\em {Proceedings of the VLDB Endowment}}, 8(5):485--496, 2015.

\bibitem{DBLP:conf/vldb/Yannakakis81}
M.~Yannakakis.
\newblock Algorithms for acyclic database schemes.
\newblock In {\em International Conference on Very Large Data Bases (VLDB)},
  pages 82--94. {IEEE}, 1981.

\bibitem{DBLP:conf/icml/ZemelWSPD13}
R.~S. Zemel, Y.~Wu, K.~Swersky, T.~Pitassi, and C.~Dwork.
\newblock Learning fair representations.
\newblock In {\em International Conference on Machine Learning (ICML)}, pages
  325--333. JMLR.org, 2013.

\bibitem{DBLP:dblp_conf/sigmod/ZhangKR14}
C.~Zhang, A.~Kumar, and C.~R{\'e}.
\newblock Materialization optimizations for feature selection workloads.
\newblock In {\em International Conference on Management of Data (SIGMOD)},
  pages 265--276. ACM, 2014.

\bibitem{DBLP:journals/corr/Zliobaite15a}
I.~Zliobaite.
\newblock A survey on measuring indirect discrimination in machine learning.
\newblock {\em CoRR}, abs/1511.00148, 2015.

\end{thebibliography}

\end{document}